\documentclass[aps,floats,twocolumn,superscriptaddress,nobalancelastpage,prb,showpacs]{revtex4}
\usepackage[final]{graphicx}
\DeclareGraphicsExtensions{.eps,.ps,.pdf}
\usepackage{amsmath,amssymb,bbold,bm}

\begin{document}


\title{
Lifshitz critical point in the cuprate superconductor YBa$_2$Cu$_3$O$_y$ from~high-field~Hall~effect~measurements 
}


\author{David LeBoeuf}
\altaffiliation{Current address : Laboratoire National des Champs Magn\'etiques Intenses, UPR 3228, (CNRS, INSA, UJF, UPS), Toulouse 31400, France}
\affiliation{D\'epartement de physique and RQMP, Universit\'e de Sherbrooke, Sherbrooke, Qu\'ebec J1K 2R1, Canada}

\author{Nicolas Doiron-Leyraud}
\affiliation{D\'epartement de physique and RQMP, Universit\'e de Sherbrooke, Sherbrooke, Qu\'ebec J1K 2R1, Canada}

\author{B. Vignolle}
\affiliation{Laboratoire National des Champs Magn\'etiques Intenses, UPR 3228, (CNRS, INSA, UJF, UPS), Toulouse 31400, France}

\author{Mike Sutherland}
\altaffiliation{Current address: Cavendish Laboratory, University of Cambridge, Cambridge CB3 OHE, UK}
\affiliation{Department of Physics, University of Toronto, Toronto, Ontario M5S 1A7, Canada}

\author{B. J. Ramshaw}
\affiliation{Department of Physics and Astronomy, University of British Columbia,
Vancouver, BC V6T 1Z1, Canada}

\author{J.~Levallois}
\altaffiliation{Current address: DPMC - Universit\'e de Gen\`eve, CH 1211, Geneva, Switzerland}
\affiliation{Laboratoire National des Champs Magn\'etiques Intenses, UPR 3228, (CNRS, INSA, UJF, UPS), Toulouse 31400, France}

\author{R. Daou}
\altaffiliation{Current address: Max Planck Institute for Chemical Physics of Solids, 01187 Dresden, Germany}
\affiliation{D\'epartement de physique and RQMP, Universit\'e de Sherbrooke, Sherbrooke, Qu\'ebec J1K 2R1, Canada}

\author{Francis Lalibert\'e}
\affiliation{D\'epartement de physique and RQMP, Universit\'e de Sherbrooke, Sherbrooke, Qu\'ebec J1K 2R1, Canada}

\author{Olivier Cyr-Choini\`ere}
\affiliation{D\'epartement de physique and RQMP, Universit\'e de Sherbrooke, Sherbrooke, Qu\'ebec J1K 2R1, Canada}

\author{Johan Chang}
\affiliation{D\'epartement de physique and RQMP, Universit\'e de Sherbrooke, Sherbrooke, Qu\'ebec J1K 2R1, Canada}

\author{Y. J. Jo}
\affiliation{National High Magnetic Field Laboratory, Florida State University, Tallahassee, Florida 32306, USA}

\author{L.~Balicas}
\affiliation{National High Magnetic Field Laboratory, Florida State University, Tallahassee, Florida 32306, USA}

\author{Ruixing Liang}
\affiliation{Department of Physics and Astronomy, University of British Columbia,
Vancouver, BC V6T 1Z1, Canada}
\affiliation{Canadian Institute for Advanced Research, Toronto, Ontario M5G 1Z8, Canada}

\author{D. A. Bonn}
\affiliation{Department of Physics and Astronomy, University of British Columbia,
Vancouver, BC V6T 1Z1, Canada}
\affiliation{Canadian Institute for Advanced Research, Toronto, Ontario M5G 1Z8, Canada}

\author{W. N. Hardy}
\affiliation{Department of Physics and Astronomy, University of British Columbia,
Vancouver, BC V6T 1Z1, Canada}
\affiliation{Canadian Institute for Advanced Research, Toronto, Ontario M5G 1Z8, Canada}

\author{Cyril Proust}
\affiliation{Laboratoire National des Champs Magn\'etiques Intenses, UPR 3228, (CNRS, INSA, UJF, UPS), Toulouse 31400, France}
\affiliation{Canadian Institute for Advanced Research, Toronto, Ontario M5G 1Z8, Canada}

\author{Louis Taillefer}
\altaffiliation{E-mail: louis.taillefer@physique.usherbrooke.ca.}
\affiliation{D\'epartement de physique and RQMP, Universit\'e de Sherbrooke, Sherbrooke, Qu\'ebec J1K 2R1, Canada}
\affiliation{Canadian Institute for Advanced Research, Toronto, Ontario M5G 1Z8, Canada}

\date{\today}


\begin{abstract}

The Hall coefficient $R_{\rm H}$ of the cuprate superconductor YBa$_2$Cu$_3$O$_y$ was measured in magnetic fields up to 60~T for a hole concentration $p$ from 0.078 to 0.152, in the underdoped regime. In fields large enough to suppress superconductivity, $R_{\rm H}(T)$ is seen to go from positive at high temperature to negative at low temperature, for $p > 0.08$. This change of sign is attributed to the emergence of an electron pocket in the Fermi surface at low temperature. At $p < 0.08$, the normal-state $R_{\rm H}(T)$ remains positive at all temperatures, increasing monotonically as $T \to 0$. We attribute the change of behaviour across $p = 0.08$ to a Lifshitz transition, namely a change in Fermi-surface topology occurring at a critical concentration $p_{\rm L} = 0.08$, where the electron pocket vanishes. The loss of the high-mobility electron pocket across $p_{\rm L}$ coincides with a ten-fold drop in the conductivity at low temperature, revealed in measurements of the electrical resistivity $\rho$ at high fields, showing that the so-called metal-insulator crossover of cuprates is in fact driven by a Lifshitz transition.
It also coincides with a jump in the in-plane anisotropy of $\rho$, showing that without its electron pocket the Fermi surface must have strong two-fold in-plane anisotropy. 
These findings are consistent with a Fermi-surface reconstruction caused by a unidirectional spin-density wave or stripe order.  

\end{abstract}

\maketitle


\section{Introduction}

The discovery of quantum oscillations in cuprates, first detected in the resistance of YBa$_2$Cu$_3$O$_y$ (YBCO)
at a hole concentration (doping) $p = 0.10$, \cite{Doiron:Nat07} revealed the presence of a small closed pocket in the Fermi surface of these materials at low temperature, 
deep in the underdoped region of the phase diagram. Observed in magnetic fields large enough to suppress superconductivity, 
the oscillations were soon also detected in the magnetization,\cite{Jaudet:PRL08,Sebastian:Nat08} in all cases giving the same dominant frequency, 
corresponding to 2~\% of the Brillouin zone area. 
Oscillations of a similar frequency were also observed in the stoichiometric cuprate YBa$_2$Cu$_4$O$_8$,\cite{Yelland:PRL08,Bangura:PRL08} 
whose doping $p \simeq 0.14$.


The fact that the oscillations are observed in a metallic state characterized by a negative Hall coefficient indicated that the associated Fermi pocket is electron-like.\cite{moi:Nat07} 
%
This implies a qualitative change in the Fermi surface of hole-doped cuprates as a function of doping, from the large hole-like cylinder observed 
in the overdoped regime,\cite{Mackenzie1996,Hussey2003,Plate2005,Vignolle:Nat08}
at $p \simeq 0.25$, to the small electron-like pocket seen in the underdoped regime. 
There must therefore be a quantum critical point that separates these two distinct metallic states (at $T = 0$, in the absence of superconductivity), 
at a doping $p^\star$ somewhere between $p \simeq 0.25$ and $p \simeq 0.14$. 
The standard mechanism for producing a small electron pocket out of a large hole surface is a reconstruction caused by the onset of a new periodicity 
that breaks the translational symmetry of the lattice,\cite{Chakravarty2008} 
as in the case of a density-wave order,\cite{Chubukov1997,Lin2005,Millis:PRB07,Chakravarty:PNAS08,Harrison:PRL09} 
but scenarios without broken translational symmetry also exist.\cite{Yang2006}

In this Article, we shed light on the nature and origin of the Fermi-surface transformation by studying the temperature and doping evolution 
of the Hall coefficient and electrical resistivity of underdoped YBCO below $p^\star$. 
Our main finding is the disappearance of the electron pocket as the doping is reduced below a critical doping $p_{\rm L} = 0.08$. 
This change in Fermi-surface topology, called a Lifshitz transition, marks a second $T = 0$ critical point in the phase diagram of YBCO
(distinct from the critical point at $p^\star$). 
We show that the loss of the high-mobility electron pocket coincides with a dramatic rise in the low-temperature resistivity, 
thereby elucidating the enigmatic metal-insulator crossover of cuprates.\cite{Ando1995,Boebinger1996} 
We show that the Lifshitz transition also coincides with an increase in the in-plane anisotropy of the resistivity,\cite{Ando:PRL02}
evidence that the remaining Fermi surface must have a strong two-fold in-plane anisotropy.
This points to a unidirectional density-wave order as the underlying cause of reconstruction, 
consistent with the anisotropic incommensurate spin-density-wave order observed by neutron scattering at low temperature for
$p < 0.08$.\cite{Hinkov2008,Haug2009}

The Article is organized as follows. 
In sec.~II, we describe the samples studied and the measurement techniques used.
In sec.~III, we present the Hall data and interpret them in terms of an electron pocket in the Fermi surface, whose characteristics and consequences are examined.
In sec.~IV, we present evidence that the electron pocket vanishes below $p = 0.08$, and show how this disappearance coincides with what has been called a
`metal-insulator' transition.
In sec.~V.A, we briefly review two scenarios of Fermi-surface reconstruction: one due to commensurate antiferromagnetic order, applied to electron-doped cuprates, and one due to stripe order, applied to hole-doped cuprates such as La$_{2-x-y}$Sr$_x$Eu$_y$CuO$_4$ (Eu-LSCO). 
We then examine how the latter scenario applies to YBCO.
In sec.~V.B, we discuss the relation between Fermi-surface reconstruction and the pseudogap phase.
In sec.~V.C, we show how superconductivity is weakened in the region of doping where the electron pocket is present in the Fermi surface.    
We summarize in sec.~VI.


\section{Samples and Methods}

The eight samples used for measurements of the Hall coefficient $R_{\rm H}$ are uncut, unpolished, detwinned crystals of YBa$_2$Cu$_3$O$_y$
 grown in a nonreactive BaZrO$_3$ crucible 
from high-purity starting materials.\cite{liang:PhysicaC00} 
The oxygen atoms in the CuO chains were made to order into the stable superstructure specific to the given oxygen concentration $y$.\cite{liang:PhysicaC00}
The main characteristics of those samples are listed in Table \ref{tab:hall}. 
The oxygen atoms in the sample with oxygen content $y = 6.48$ were first fully ordered in an ortho-II structure (I), then disordered (II) and then partially 
re-ordered (III), 
thereby tuning the doping within one and the same crystal without changing the oxygen content, the impurity level or the dimensions.
The two samples used for measurements of the in-plane electrical resistivity $\rho_a$ are uncut, unpolished, detwinned crystals of YBCO grown in
yttria-stabilized zirconia crucibles. They had $T_c = 47$~K and 62~K, giving $p = 0.08$ and 0.11, respectively.
In all cases, the electric current was applied along the $a$ axis of the orthorhombic structure, perpendicular to the CuO chains which run along the $b$ axis. 
The hole concentration (doping) $p$ of all ten samples was determined from the superconducting transition temperature $T_c$,\cite{Liang:PRB06}
defined as the temperature where the resistance goes to zero. The uncertainty on $T_c$ is $\pm 0.2$~K.
Typical sample dimensions are (20-50 $\mu$m) $\times$ (500-800 $\mu$m) $\times$ (500-1000 $\mu$m) [(thickness) $\times$ (width) $\times$ (length)].
Transport properties were measured via gold evaporated contacts in a six-contact geometry for the Hall resistance and 
diffused silver epoxy contacts in a four-contact geometry for the resistivity. 
%
%
The magnetic field $B$ was applied along the $c$ axis of the orthorhombic structure, perpendicular to the CuO$_2$ planes.
The samples with $y = 6.67$ and $y = 6.75$ were measured in a steady magnetic field up to 45 T in the hybrid magnet at the NHMFL in Tallahasee. 
All the other samples were measured in a pulsed magnetic field up to 50-60 T in a resistive magnet at the LNCMI in Toulouse.
The sample with $y=6.75$ was measured in both labs, yielding identical results.

\begin{table}[h]
	\begin{center}
		\begin{tabular}{lcc}
\hline

$~~y$ & $T_{\rm c}$ &    $p$ \\

\hline

      &    (K)      &  (holes/Cu)  \\

\hline
\hline

6.45     & 44.5   & 0.078   \\
6.48 II	 & 49.5 & 0.083   \\
6.48 III & 51.0   & 0.085   \\
6.48 I	 & 53.0   & 0.088   \\
6.51		 & 57.3   & 0.097   \\
6.54		 & 61.3   & 0.108   \\
6.67		 & 66.0   & 0.120   \\
6.75		 & 74.8   & 0.132   \\
6.80     & 77.9   & 0.135   \\
6.86     & 91.1   & 0.152   \\


\hline
\hline

	\end{tabular}
	\end{center}
\caption{Main characteristics of the YBCO samples used for Hall effect measurements: 
the oxygen content $y$; 
the superconducting transition temperature in zero magnetic field, $T_c$;
the hole concentration (doping) $p$, obtained from $T_c$.\cite{Liang:PRB06}
}
\label{tab:hall}
\end{table}


\begin{figure}[t]
\center
	\includegraphics[width=8cm]{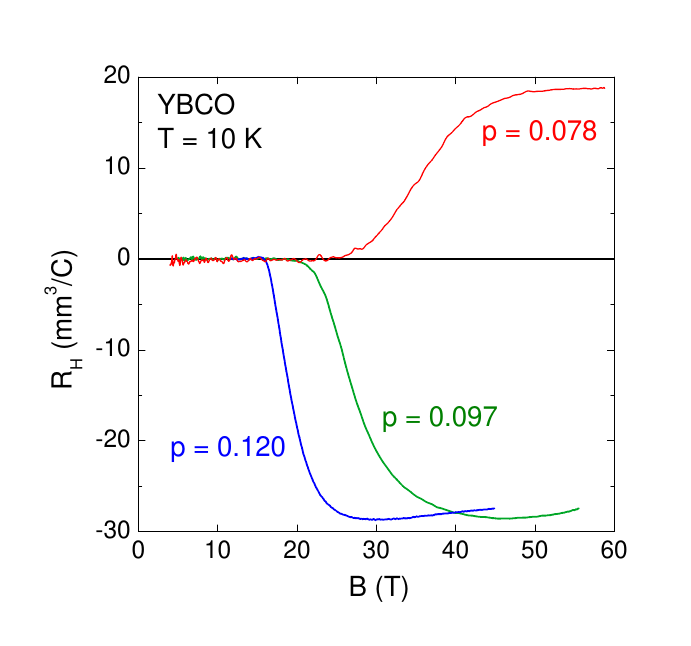}
	\caption{
	Hall coefficient $R_{\rm H}$ of YBCO at $T = 10$~K as a function of magnetic field for three samples, with dopings $p$ as indicated.
	$R_{\rm H}$ is negative for $p=0.120$ and $p=0.097$, as previously reported. \cite{moi:Nat07}
	By contrast, $R_{\rm H}$ is positive for $p=0.078$,  
  pointing to a qualitative change in the Fermi surface across $p \simeq 0.08$.}
	\label{fig:RHvsB}
\end{figure}



\section{Electron pocket in the Fermi surface}

\subsection{Sign change in the Hall coefficient}

\begin{figure}[t]
\center
	\includegraphics[width=8cm]{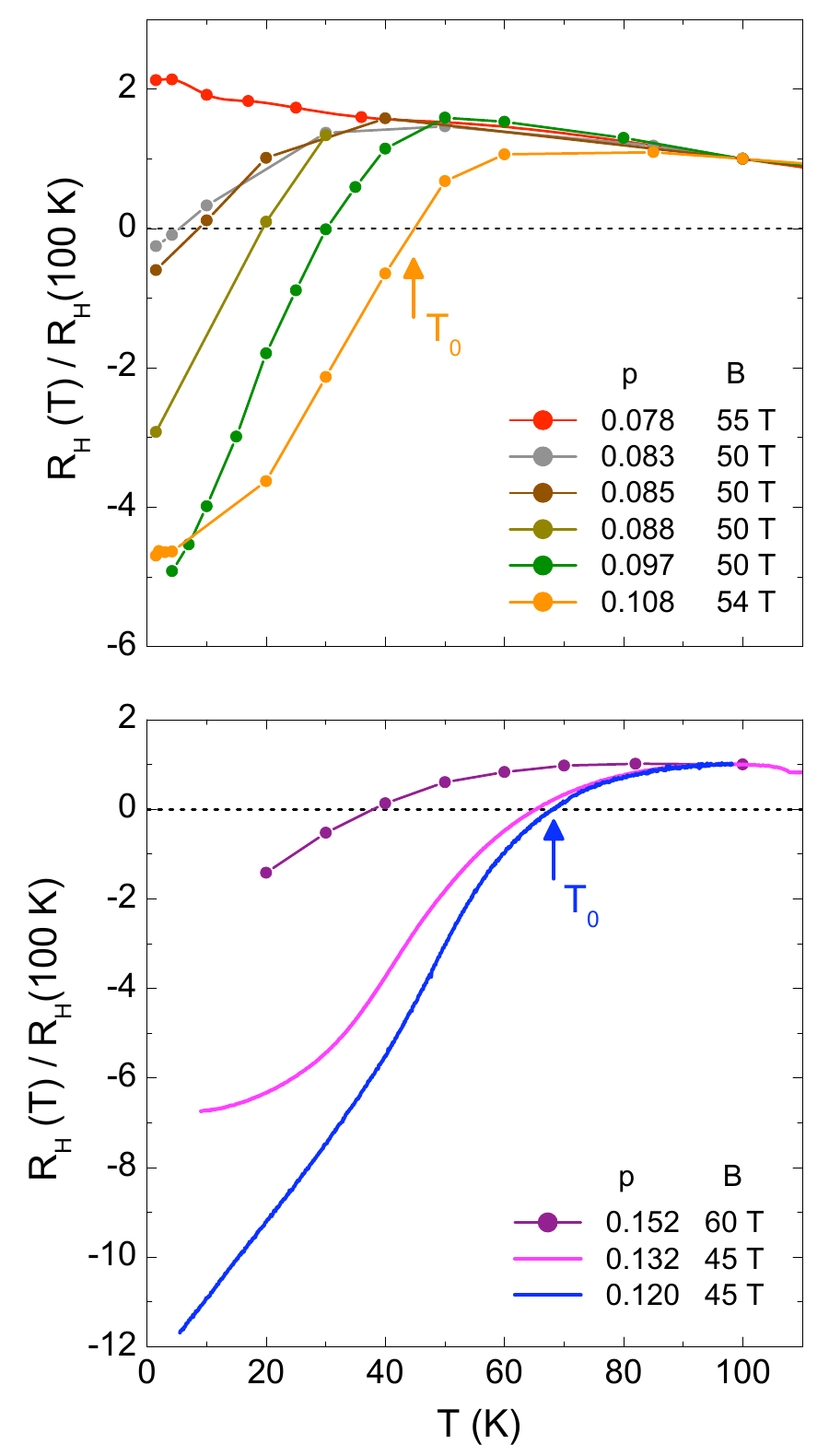}
	\caption{
	Hall coefficient of YBCO at nine different dopings $p$ as indicated, normalized to its value at $T=100$ K.
	The data is taken at the highest magnetic field $B$ reached, as indicated.
	Top panel: $p < 0.11$. Bottom panel: $p \geq 0.12$. 
	$T_0$ marks the temperature at which $R_{\rm H}(T)$ changes sign.}
	\label{fig:RH2pannel}
\end{figure}

In Fig.~\ref{fig:RHvsB}, the Hall coefficient of YBCO at $T = 10$~K is plotted as a function of magnetic field $B$ for three dopings: 
$p = 0.120$, 0.097 and 0.078. 
The flux-flow regime (or vortex-liquid phase) starts approximately at $B = 15$, 20 and 25~T, and ends approximately at $B = 30$, 45 and 50~T, respectively.
%
In the normal state beyond the flux-flow regime, $R_{\rm H}$ is seen to be positive at $p = 0.078$ but negative at $p = 0.097$ and 0.120, 
pointing to a qualitative change in the Fermi surface occuring between $p \simeq 0.08$ and $p \simeq 0.10$.
In Fig.~\ref{fig:RH2pannel}, the normal-state Hall coefficient, measured at the highest field (between 45 and 60 T), is plotted as a function of temperature for 
several dopings. We see that $R_{\rm H}(T)$ goes from positive at $T = 100$~K to negative as $T \to 0$,
except for $p = 0.078$ where $R_{\rm H}(T)$ never changes sign and simply increases monotonically with decreasing temperature. 
%

As illustrated in Fig.~3 for the case of $p = 0.12$, the temperature dependence of $R_{\rm H}(T)$ can be described using three characteristic temperatures:
1) $T_0$, the temperature at which $R_{\rm H}(T)$ changes sign;
2) $T_{\rm max}$, the temperature at which $R_{\rm H}(T)$ attains its maximal value;
3) $T_{\rm H}$, the temperature below which $R_{\rm H}(T)$ starts to show downward curvature (inflexion point).
In Fig.~4, we plot these three characteristic temperatures vs doping. 
%


We see that $T_{\rm max}$ and $T_{0}$ both peak at $p \simeq 1/8$. 
Below $p = 0.12$, $T_{0}$ decreases monotonically (and linearly) to zero as $p \to 0.08$, whereas 
$T_{\rm max}$ remains finite down to $p = 0.083$, and then drops suddenly to zero when the doping decreases slightly, to $p = 0.078$.
By contrast, the onset of the downward curvature, $T_{\rm H}$, is relatively flat with doping, decreasing monotonically from 
$\simeq 160$~K at $p = 0.10$ to $\simeq 120$~K at $p = 0.14$.
Note that $T_{\rm H}$ is determined from published curves of $R_{\rm H}(T)$,\cite{Segawa2004}
measured in low fields above $T_c$;
at two dopings, we compare these with data taken in two of our samples, in low fields up to 200~K (Fig.~3).

To summarize, we observe a drop in $R_{\rm H}(T)$ at all dopings from $p = 0.083$ to $p = 0.152$, inclusively. 
At $p = 0.078$, there is no drop down to the lowest temperatures.
Above $p = 0.152$, we cannot suppress superconductivity entirely with a field of 60~T.
However, measurements on a sample with $T_c = 93$~K ($y = 6.92$; not shown) allow us to put an upper bound on 
$T_{\rm max}$, such that $T_{\rm max} \leq 60$~K at $p = 0.161$.
Our findings are in good overall agreement with earlier low-field data,\cite{Ito1993,Mizuhashi1995,Abe1999,Xu1999,Segawa2004} which show a clear peak (maximum) in $R_{\rm H}(T)$ for $p$ ranging from 0.09 to 0.14, and no peak (above $T_c$) for $p \leq 0.08$ or $p \geq 0.16$. 


\begin{figure}[t]
\center
	\includegraphics[width=8cm]{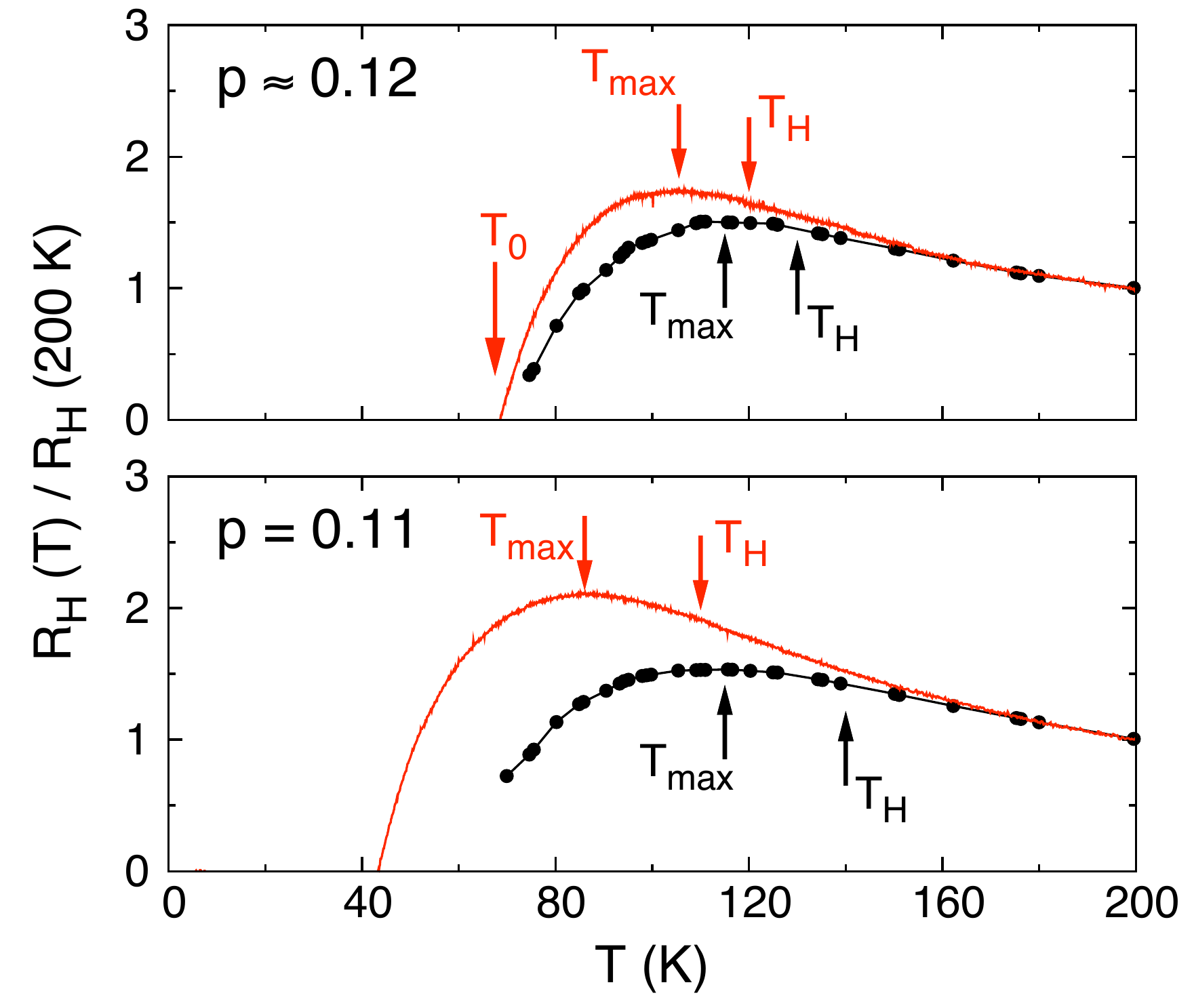}
		\caption{
		Temperature dependence of $R_{\rm H}$, normalized to its value at 200~K, in YBCO at $p \simeq 0.12$ (top) and $p = 0.11$ (bottom).
		In order to have higher resolution and continuous temperature sweeps, data was taken on two of our samples ($y=6.67$ and $y=6.54$) 
		in low field (15~T) up to 200~K. 
		Top panel: the continuous (red) curve is our data on a sample with $T_c = 66$~K ($p = 0.12$) at $B = 15$~T; 
		the black dots are the data of ref.~\onlinecite{Segawa2004} for a sample (labelled $y=6.80$) with $T_c = 69$~K ($p = 0.125$) at $B = 14$~T.
		The arrows indicate the position of the three characteristic temperatures that describe the temperature evolution of $R_{\rm H}(T)$:
		1) $T_0$, the temperature at which $R_{\rm H}(T)$ changes sign;
		2) $T_{\rm max}$, the temperature at which $R_{\rm H}(T)$ attains its maximal value;
		3) $T_{\rm H}$, the temperature below which $R_{\rm H}(T)$ starts to show downward curvature.
		Bottom panel: the continuous (red) curve is our data on a sample with $T_c = 60$~K ($p = 0.11$) at $B = 15$~T; 
		the black dots are the data of ref.~\onlinecite{Segawa2004} for a sample (labelled $y=6.70$) with $T_c = 60$~K ($p = 0.11$) at $B = 14$~T.
		$T_{\rm max}$ and $T_{\rm H}$ are defined as for the top panel.
    }
	\label{fig:RHdiag}
\end{figure}



\begin{figure}[t]
\center
	\includegraphics[width=8cm]{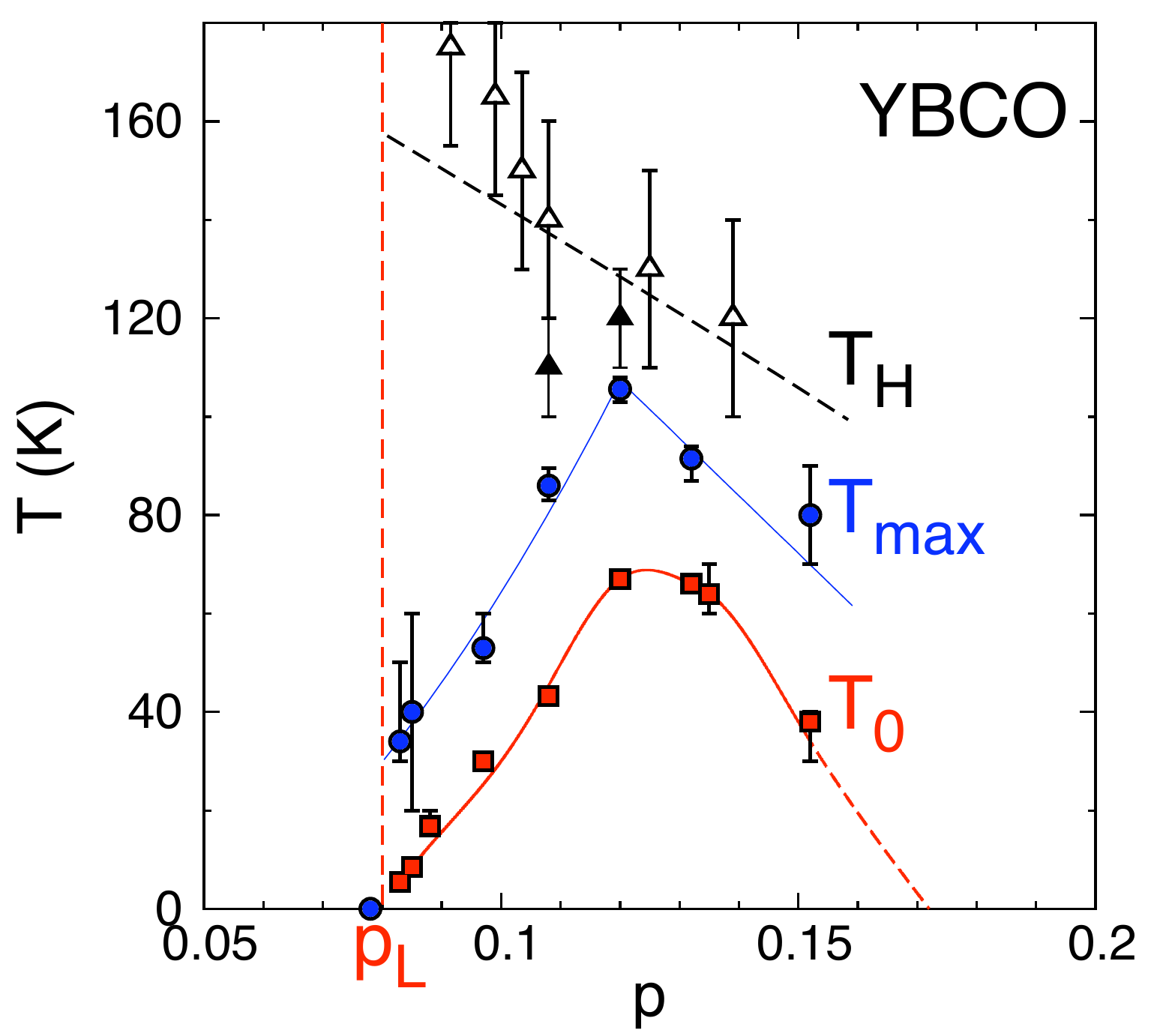}
		\caption{
		Doping dependence of the three characteristic temperatures defined in Fig.~3: 
		$T_0$ (red squares),
		$T_{\rm max}$ (blue circles), and
		$T_{\rm H}$ (triangles).
		These temperatures describe the evolution of $R_{\rm H}(T)$ as it drops to negative values upon cooling. 
		$T_{\rm H}$ is where the downward deviation starts; it may be thought of as the onset of the
		Fermi-surface reconstruction that leads to the formation of the electron pocket.
		Our data (full triangles) are compared to published data (open triangles, from ref.~\onlinecite{Segawa2004}).
		The vertical (red) dashed line marks the critical doping $p \equiv p_{\rm L} = 0.08$ below which no drop in $R_{\rm H}(T)$ is
		observed down to the lowest temperature.
		All other lines are a guide to the eye.  
		}
   \label{fig:T_0}
\end{figure}



\subsection{Evidence for an electron pocket}

A number of experimental observations combine to show that the negative $R_{\rm H}$ in YBCO is due to 
a small, closed, high-mobility electron pocket in the Fermi surface.
Earlier measurements \cite{Ong:Hall-YBCO} of $R_{\rm H}$ in a YBCO sample with $T_c \simeq 60$~K ($p \simeq 0.11$) 
in fields up to 23~T revealed a sign change consistent with subsequent measurements in higher fields, \cite{moi:Nat07} 
but they were interpreted at the time as coming from a negative flux-flow contribution to the Hall signal due to moving vortices.\cite{Ong:Hall-YBCO}
Two facts now rule out this flux-flow interpretation.
First, flux flow yields a contribution to the Hall resistance which is strongly non-linear in $B$, 
so that the Hall coefficient should depend strongly on magnetic field, and hence so should the sign-change temperature $T_0$.
In YBCO at $p = 0.12$, this is not the case: $T_0$ is entirely independent of $B$ at all fields,\cite{moi:Nat07}
as shown in the $B$-$T$ phase diagram of Fig.~5.
Secondly, the movement of vortices is readily detectable in the Nernst effect, to which it makes a large positive contribution.
It has recently been shown that in YBCO at $p = 0.12$ the vortex Nernst signal at $T > 9$~K is negligible for fields greater than 30~T,\cite{Chang:PRL10}
proving that the large negative $R_{\rm H}$ at low temperature seen at that doping does not come from flux flow.
This is illustrated in Fig.~6, where we see that $R_{\rm H}$ vs $B$ at $T = 10$~K does indeed saturate above 30~T.
So the negative $R_{\rm H}$ in YBCO is not a consequence of flux flow, but a property of the Fermi surface.


\begin{figure}[t]
\center
	\includegraphics[width=8cm]{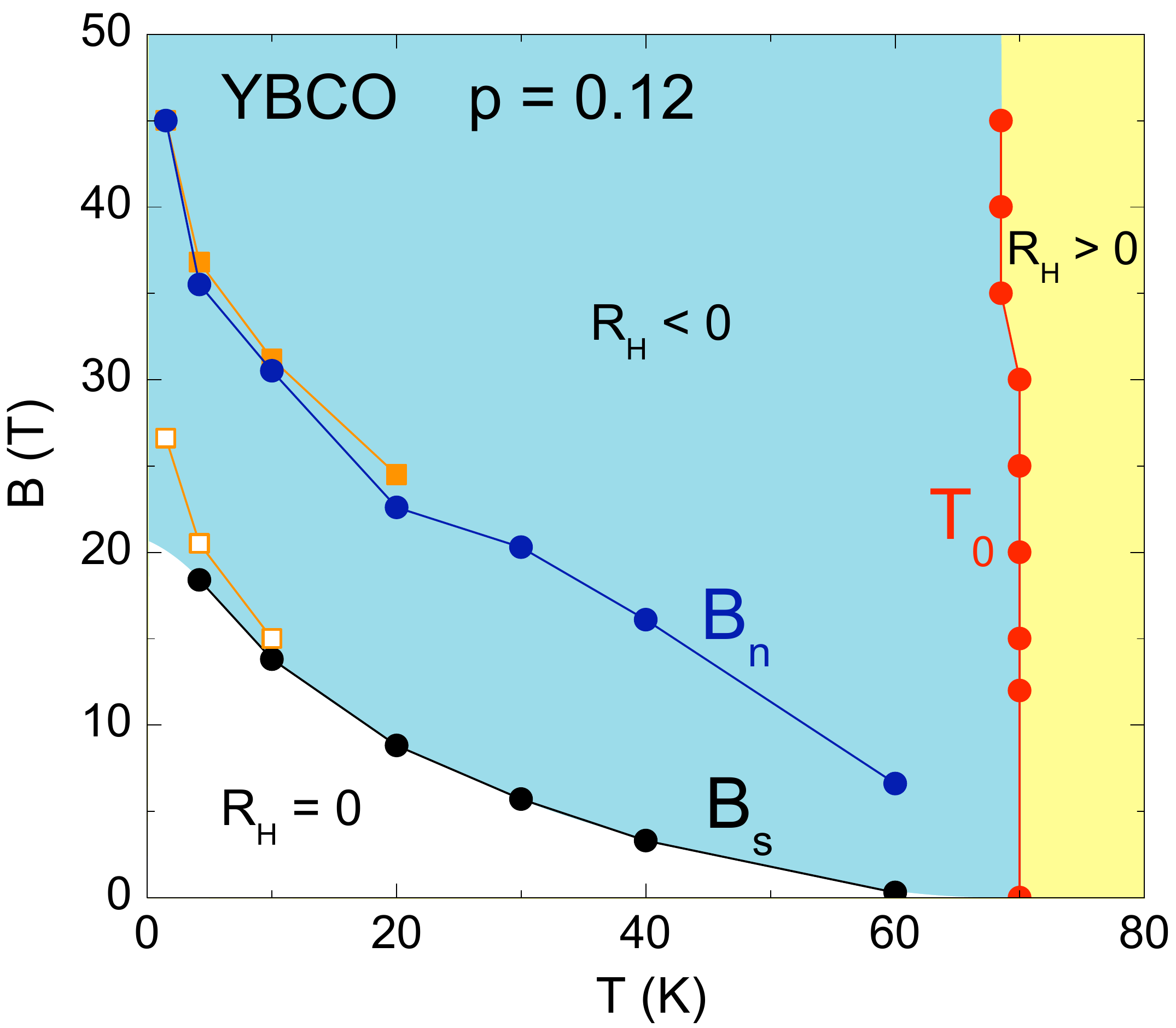}
		\caption{
		Field-temperature phase diagram of YBCO for $p = 0.12$, obtained from transport measurements (see ref.~\onlinecite{note}).
		The solid vortex phase (where $R_{\rm H} = \rho = 0$) ends at $B_{\rm s}(T)$ and the normal state is reached above 
		$B_{\rm n}(T)$, in the sense that flux-flow contributions to transport have become negligible.   
		$B_{\rm s}$ and $B_{\rm n}$ are defined from resistance measurements (orange squares from $R_{\rm xy}$ and solid circles from $R_{\rm xx}$;
		see ref.~\onlinecite{moi:Nat07}).
		$T_0$ is the temperature below which $R_{\rm H}$ changes sign from positive to negative.
		All lines are a guide to the eye.  
		}
   \label{fig:B-T}
\end{figure}



\begin{figure}[t]
\center
	\includegraphics[width=8cm]{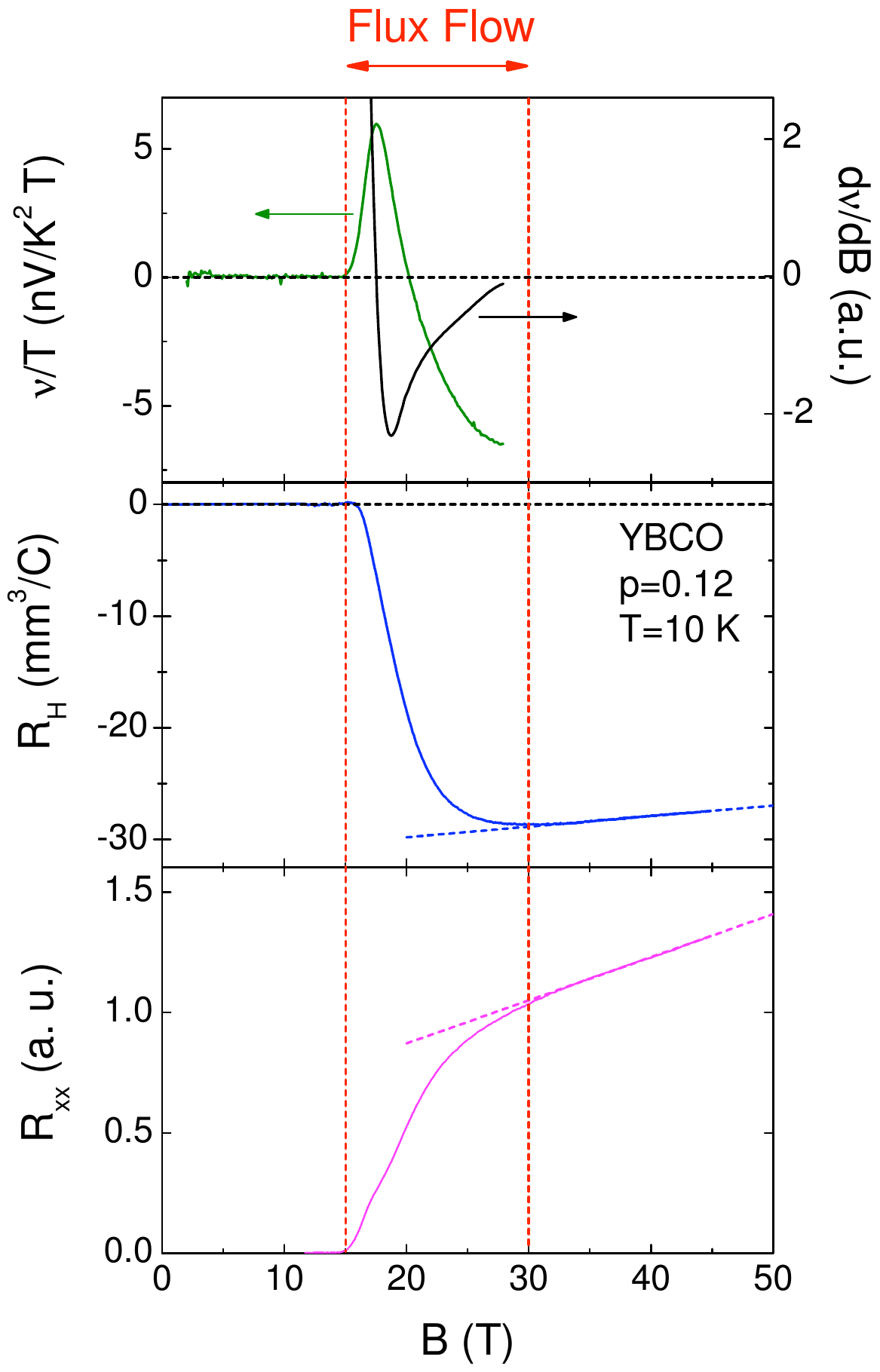}
	\caption{
	Field dependence of the Nernst coefficient (top panel; from ref.~\onlinecite{Chang:PRL10}; $T = 9$~K), 
	the Hall coefficient (middle panel; from ref.~\onlinecite{moi:Nat07})
	and the in-plane resistivity (bottom panel; from ref.~\onlinecite{moi:Nat07})
	of YBCO at $p = 0.12$ and $T = 10$~K. 
	Above a threshold field of $\sim 30$~T 
	(indicated by the vertical dotted red line on the right), 
	the Nernst coefficient $\nu/T$ (top panel; green curve)
	saturates to its negative quasiparticle value, as demonstrated by its derivative 
	(top panel; black curve) which goes to zero as $B \to 30$~T. 
	This saturation shows that the positive contribution to the Nernst coefficient from 
	superconducting fluctuations has become negligible above $\sim 30$~T (for $T > 9$~K).
	Above this field, the Hall coefficient is almost flat (dashed blue line in middle panel)
	and the in-plane resistivity shows a linear magnetoresistance (dashed magenta line in bottom panel).
	We conclude that above $\sim 30$~T the flux-flow contribution to the transport properties 
	of YBCO (at $p = 0.12$) is negligible, 
	and the transport coefficients at high field are purely a property of the normal state.
	}
	\label{fig:Flux-flow}
\end{figure}


The Fermi surface of YBCO at $p = 0.10$-0.11 is known to contain two small closed pockets (cylinders), 
responsible for the quantum oscillations seen below 10~K and above $\sim 30$~T.
\cite{Doiron:Nat07,Jaudet:PRL08,Sebastian:Nat08,Audouard2009,Ramshaw2010}
The fact that quantum oscillations in the longitudinal resistance $R_{xx}$ are out of phase with the oscillations in the 
transverse (Hall) resistance $R_{xy}$ \cite{Jaudet2009} is direct evidence that the oscillations are
due to electron-like carriers.\cite{Coleridge1989}
The two oscillatory components have very similar frequencies (not resolved in the first measurements\cite{Doiron:Nat07}), 
near $F \simeq 500$~T,\cite{Audouard2009,Ramshaw2010}
corresponding to a carrier density $n = F / \Phi_0 \simeq 0.04$~per planar Cu atom.\cite{Doiron:Nat07}
They also have similar mobilities (or Dingle temperatures).\cite{Ramshaw2010}
The pair of Fermi cylinders has been associated with the pair of CuO$_2$ planes (or bi-layer) in the unit cell of YBCO,
with bi-layer coupling lifting the two-fold degeneracy and causing the two frequencies to be slightly split.\cite{Audouard2009}
Each of these two pockets is expected to make a comparable contribution to the Hall effect, 
of magnitude $|R_{\rm H}| = 1/en \simeq 30$~mm$^3$/C, assuming they are isotropic in the plane (circular). 
If both are electron-like, then $R_{\rm H} \simeq - 1/2en \simeq - 15$~mm$^3$/C. 
This is roughly half the measured value of $R_{\rm H} \simeq- 28$~mm$^3$/C at $p = 0.097$ (see Fig.~1).  
The larger measured value may be evidence that the electron pocket is not circular but squarish, 
with angles whose higher curvature would enhance $|R_{\rm H}|$ if scattering is anisotropic.\cite{Ong1991}

The Seebeck coefficient $S$ (or thermopower) provides a separate measure of the sign of charge carriers.
Recent measurements in YBCO at $p = 0.12$ up to 28~T yield 
$S/T = -~0.4~\mu$V K$^{-2}$, in the normal state as $T \to 0$.\cite{Chang:PRL10}
The negative sign confirms that the dominant carriers, {\it i.e.} those with the highest mobility, are electron-like.

The magnitude of $S$ in metals can be estimated approximately from the following expression, valid as $T \to 0$:\cite{Behnia2004}

\begin{equation}
\frac{S}{T} \simeq \pm \frac{\pi^2}{2} \frac{k_{\rm B}}{e} \frac{1}{T_{\rm F}}
\end{equation} 

where $k_{\rm B}$ is Boltzmann's constant, $e$ is the electron charge, and $T_{\rm F}$ the Fermi temperature.    
The sign of $S$ is controlled by the carrier type: positive for holes, negative for electrons.
Eq.~1 can be used to check if the pocket seen in quantum oscillations is small enough to account for the large 
measured value of $S/T$. 
From the oscillation frequency $F = 540 \pm 4$~T and cyclotron mass $m^\star = 1.76 \pm 0.07~m_0$,\cite{Jaudet:PRL08} 
where $m_0$ is the bare electron mass, we obtain 
$T_{\rm F} = (e \hbar / k_{\rm B}) (F / m^\star) = 410 \pm 20$~K.  
From Eq.~1, this small value of $T_{\rm F}$ yields $| S/T | = 1.0~\mu$V K$^{-2}$.
This shows that the Fermi pocket measured by quantum oscillations has a sufficiently small Fermi energy to account for the large negative
thermopower at $T \to 0$, if it is an electron pocket.

We conclude that quantum oscillations and the Hall and Seebeck coefficients measured in YBCO near $p \simeq 0.1$ 
are quantitatively consistent with a small closed electron pocket in the Fermi surface (split into two pockets by bi-layer coupling).
We emphasize that the electron-pocket state is not a field-induced state.
Although previous authors have shown and stated that the drop in $R_{\rm H}(T)$ below $T_{\rm max}$ in YBCO is independent of $B$,\cite{Segawa2004}
it is worth stressing this point again.
It has been shown that, for $p=0.12$,  $R_{\rm H}(T)$ crosses from positive to negative at the same temperature $T_0 = 70$~K for all fields (as plotted in Fig.~5)(see Fig. S1b in the Supplementary Material of ref.\cite{moi:Nat07}).
Moreover, the effect of the field is simply to push superconductivity down to lower temperature.
As $B$ increases, a single $B$-independent $R_{\rm H}(T)$ curve is revealed.
This proves that the field does not induce Fermi-surface reconstruction but it simply allows us to reveal it at temperatures below $T_c$.
In other words, the electron-pocket state at low temperature is a `field-revealed' state.





\subsection{Location of the electron pocket}

Because $c$-axis transport in cuprates is dominated by states that are located near the $(\pi, 0)$ point (and equivalent points) in the Brillouin zone, \cite{Chakravarty1993}
measurements of the resistivity $\rho_c$ along the $c$-axis of the YBCO crystal structure provides information on the $k$-space location of 
the conducting states on the Fermi surface.
Recent measurements of $c$-axis transport on YBCO crystals with $p \simeq 0.11$ reveal large quantum oscillations in $\rho_c$ vs $B$.\cite{Ramshaw2010}
These oscillations arise from the same in-plane orbits that produce the quantum oscillations seen in either in-plane transport\cite{Doiron:Nat07} or magnetization.\cite{Jaudet:PRL08,Sebastian:Nat08} 
Their amplitude is up to 50 \% of the background resistance.\cite{Ramshaw2010}  
Such large oscillations can only come from a Fermi surface pocket located at $(\pi, 0)$, for it is those pockets which dominate the $c$-axis transport. 
We conclude that the small closed high-mobility electron pocket in the Fermi surface of YBCO is indeed located at this point in the Brillouin zone. 
(Note that the two split frequencies (near 500~T) are both detected in $\rho_c$, evidence that both are located near $(\pi, 0)$.)
This is consistent with scenarios that attribute the electron pocket to a reconstruction of the Fermi surface caused by an ordered state that breaks 
the translational symmetry of the lattice,\cite{Chakravarty2008}
as these invariably yield an electron pocket at $(\pi, 0$).
This includes the stripe scenario\cite{Millis:PRB07} discussed in section V.A. 

A different interpretation of the quantum oscillations measured in YBCO has recently been proposed 
in terms of a three-pocket scenario which neglects bi-layer splitting and attributes
the two closeby frequencies near $F \simeq 500$~T to a small electron pocket at $(\pi,0)$ and a small hole pocket at $(\pi/2,\pi/2)$,
plus an additional closed large hole pocket.\cite{Sebastian2010b}
It is difficult to see how this scenario can be reconciled with the large negative Hall and Seebeck coefficients measured in YBCO at the same doping.
Given their similar carrier densities and mobilities,\cite{Ramshaw2010} 
the two small pockets would necessarily make comparable contributions to $R_{\rm H}$ and $S/T$, but of opposite sign, thereby cancelling out, 
leaving the resulting coefficients to be dominated by the larger hole pocket, giving positive values,
contrary to observation.


\subsection{Mobility of the electron pocket}

For the electron pocket to dominate the Hall and Seebeck signals, its mobility at $T \to 0$ must be much higher than that of the remaining sheets of the Fermi surface. 
An indication that this is indeed the case comes from a recent two-band analysis of the longitudinal and Hall resistances of YBa$_2$Cu$_4$O$_8$ 
as a function of magnetic field,\cite{Rourke2010} 
which are very similar to the longitudinal and Hall resistances of YBCO.\cite{moi:Nat07}
In a two-band model of one electron sheet ($e$) and one hole sheet ($h$), this analysis finds the conductivity of the electron pocket to be one order of magnitude larger 
than the hole conductivity at low temperature:  $\sigma_e \simeq 10~\sigma_h$ at $T \to 0$. 
(See Sec.~IV.B for independent evidence that in YBCO the mobility of the electron pocket is also one order of magnitude higher than that of the other carriers.)

The magnitude and sign of $R_{\rm H}(T)$ depend on the relative magnitude of the electron and hole mobilities.
A change in the impurity scattering will in general alter the mobilities.
Addition of Zn impurities at a fixed doping $p \simeq 0.11$ makes the drop in $R_{\rm H}(T)$ shallower,
and at high enough Zn concentration $R_{\rm H}(T)$ no longer extrapolates to a negative value at $T=0$.\cite{Mizuhashi1995}
By contrast, $T_{\rm H}$ is not affected by Zn doping.\cite{Mizuhashi1995,Abe1999}


The other factor that will influence the temperature dependence of $R_{\rm H}(T)$ is inelastic scattering.
The rapid rise in $R_{\rm H}(T)$ upon warming from $T=0$, for example in the $p=0.12$ data (Fig.~2), may be due to the fact that
inelastic scattering in cuprates is strongest in the $(\pi, 0)$ direction,\cite{Abdel-Jawad2006}
{\it i.e.} where the electron pocket is located. 
This could rapidly alter the balance of electron to hole mobilities in favour of a positive Hall coefficient, especially as this anti-nodal scattering 
grows linearly with temperature, as does $R_{\rm H}(T)$. 
On the other hand, the rapid rise of $R_{\rm H}(T)$ at low temperature may also come from an increase in the conductivity of the other, hole-like 
parts of the Fermi surface, as found in the two-band analysis of YBa$_2$Cu$_4$O$_8$ data.\cite{Rourke2010}

In summary, the overall temperature dependence of $R_{\rm H}(T)$ between $T_{\rm H}$ and $T=0$ is complex, and expected to be sample dependent.
This is what we find comparing our data with earlier data on YBCO. 
In Fig.~3, we see that the data of ref.~\onlinecite{Segawa2004} on a sample with $T_c = 60$~K,
for example, gives $T_{\rm max} \simeq 115$~K, while our sample with the same $T_c$ shows a peak at $T_{\rm max} \simeq 85$~K.


\subsection{Consequences of the electron pocket}



\subsubsection{Large magnetoresistance in $\rho_a$ and $\rho_c$}

In a metal, the motion of charge carriers in a transverse magnetic field yields a positive orbital magnetoresistance (MR).
This MR is small when the Fermi surface contains only one closed pocket, and zero if that pocket is isotropic.
A standard mechanism for producing a large MR is the presence in the Fermi surface of one electron pocket and one (or more)
hole pocket(s).
The analysis of the longitudinal and Hall resistances of YBa$_2$Cu$_4$O$_8$ mentioned above shows that a two-band model
describes the data in detail,\cite{Rourke2010} 
and does therefore account for the large MR in YBa$_2$Cu$_4$O$_8$.
It is then natural to attribute the large positive MR 
seen in low-temperature high-field measurements of $\rho_a$ on YBCO at $p = 0.10$-0.12,
very similar to that seen in YBa$_2$Cu$_4$O$_8$,\cite{moi:Nat07}
to the known presence of the electron pocket in that range of doping, temperature and field.

\begin{figure}[t]
\center
	\includegraphics[width=8cm]{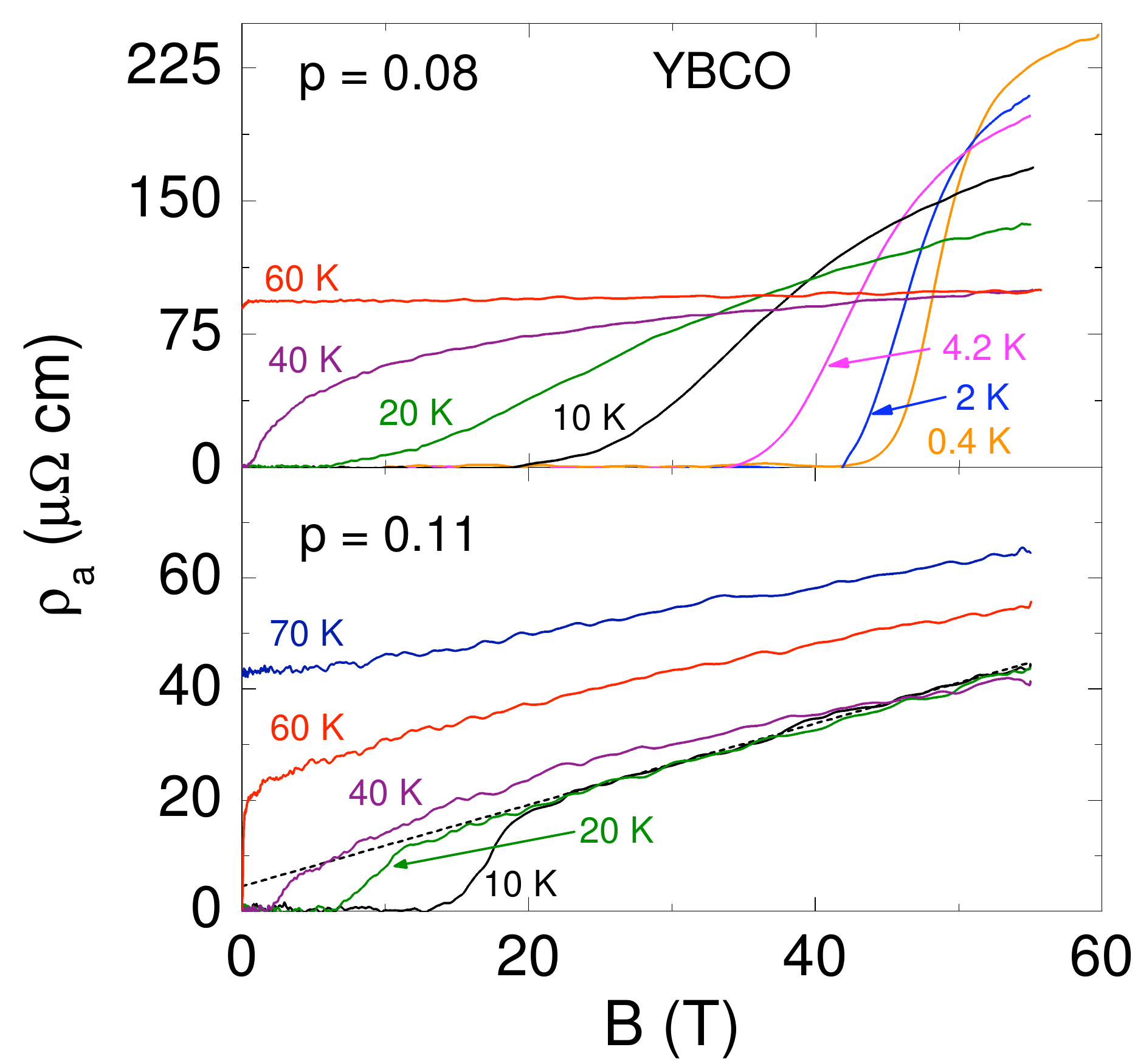}
	\caption{
	In-plane resistivity of YBCO as a function of magnetic field $B$ for two dopings, $p = 0.08$ (top) and $p = 0.11$ (bottom), 
	at temperatures as indicated.
	The isotherms at 60~K show that the magnetoresistance (MR) is weak at $p = 0.08$ and strong at $p = 0.11$.
	Relative to $\rho_a$(55~T), the strong, roughly linear positive MR at $p = 0.11$ (i.e. the slope of $\rho_a$ vs $B$ at 55~T) 
	grows with decreasing temperature.
	The dotted line in the bottom panel is a linear fit to the normal-state resistivity of the $p = 0.11$ sample at 10~K, 
	whose extrapolation to $T=0$ gives a rough estimate of the
	MR-free resistivity $\rho(0)$, of order 5 $\mu\Omega$~cm at low temperature.
	}
	\label{fig:rhovsB}
\end{figure}

In Fig.~\ref{fig:rhovsB}, we show $\rho_a$ vs $B$ for our resistivity sample with $p = 0.11$.
(Our data at $p=0.11$ is very similar to previously published data on an underdoped YBCO crystal with $T_c = 57$~K ($p = 0.1$).\cite{Rullier-Albenque2007})
Below 100~K, the MR is nearly linear and it becomes strong at low temperature:
at 10~K, $\rho_a$ doubles in going from 20~T to 55~T.
%

\begin{figure}[t]
\center
	\includegraphics[width=8cm]{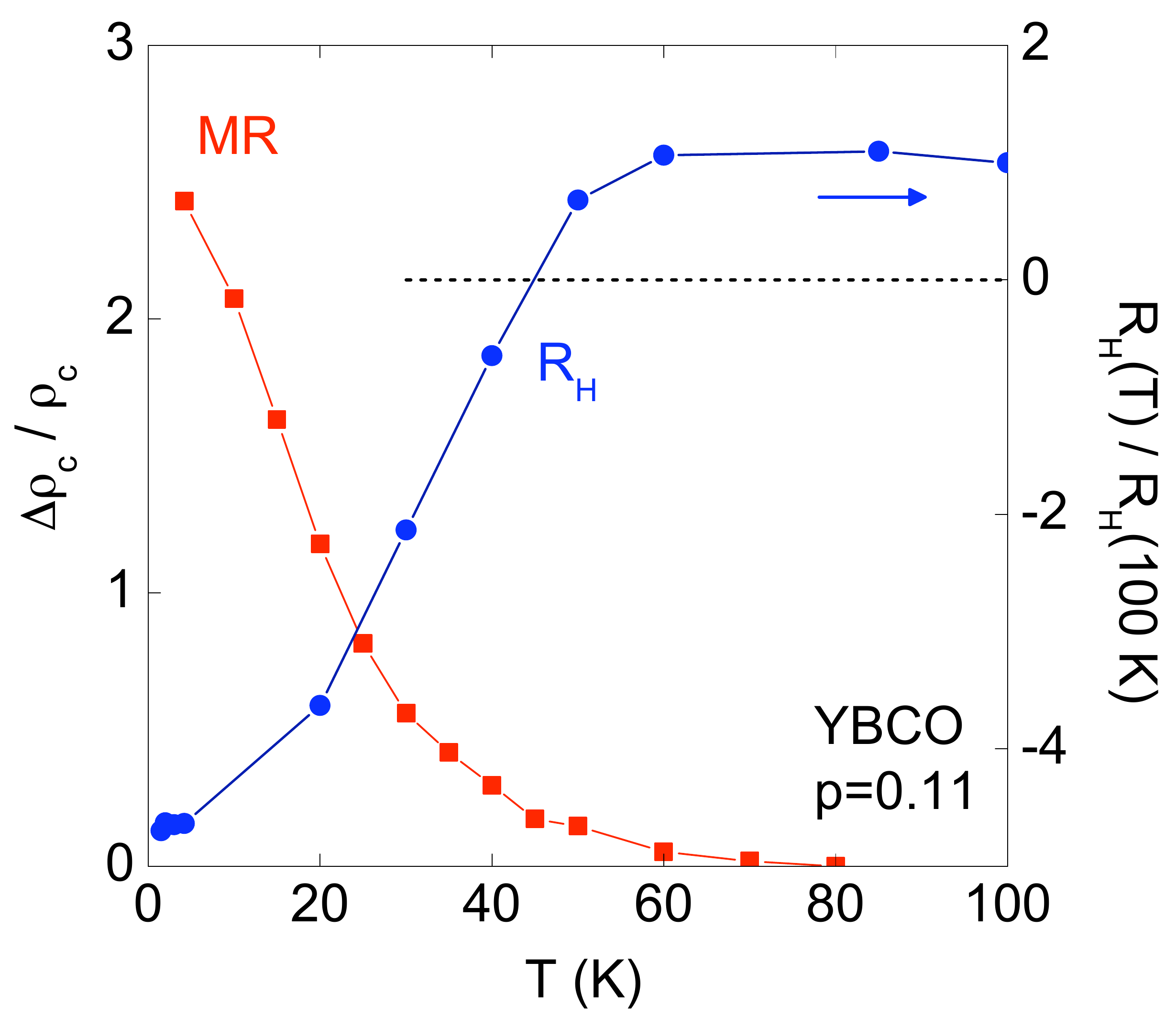}
	\caption{$c$-axis magnetoresistance (MR) of YBCO at $p=0.11$
   (red squares, left axis; from Ref.~\onlinecite{Balakirev2000},
   sample 2 with $T_c = 60$~K), plotted as $\Delta\rho_c/\rho_c$,
   where $\Delta\rho_c \equiv \rho_c(50~{\rm T}) - \rho_c(0~{\rm T})$,
   and $\rho_c(0~{\rm T})$ is obtained from a linear extrapolation of
   $\rho_c(B)$ to $B=0$.
   The in-plane Hall coefficient $R_{\rm H}(T)$ of YBCO at the same
   doping is shown for comparison (blue circles, right axis; this work,
   sample with $p=0.108$), normalized to its value at 100~K.
   The positive MR is seen to develop as the electron pocket emerges,
   as manifest in the drop of $R_{\rm H}(T)$ to negative values.	}
	\label{fig:caxisMR}
\end{figure}

The large, nearly linear MR seen in $\rho_a$ 
at $p \simeq 0.1$ (Fig.~\ref{fig:rhovsB}) is also seen in $\rho_c$.\cite{Balakirev2000}
In Fig.~\ref{fig:caxisMR}, we plot the $c$-axis MR of a YBCO sample with $T_c = 60$~K ($p = 0.11$) using data from Ref.~\onlinecite{Balakirev2000}, where $\rho_c(0)$ is obtained from a linear extrapolation of $\rho_c(B)$ to $B = 0$. The MR is strong at low temperature: 
at 10~K, $\rho_c$ nearly doubles in going from 20~T to 60~T.
Also plotted in Fig.~\ref{fig:caxisMR} is the in-plane $R_{\rm H}(T)$ of
YBCO at the same doping. We clearly observe that the MR in $\rho_c$ is negligible at 100~K and its gradual growth upon cooling tracks the drop in the in-plane $R_{\rm H}(T)$.
This close correlation between in-plane $R_{\rm H}$ and MR in $\rho_c$ is further confirmation 
that the electron pocket in the Fermi surface of YBCO is located at ($\pi$, 0),
where it controls how much of the orbital MR in the in-plane transport is reflected in the $c$-axis conduction.


\subsubsection{Large and isotropic Nernst coefficient}

Much as the high-mobility electron pocket is responsible for the large (and negative) Hall and Seebeck coefficients in the normal state of YBCO at $T \to 0$ for 
$p = 0.10$-0.12, it is also responsible for the large quasiparticle Nernst coefficient.\cite{Chang:PRL10} 

The magnitude of the Nernst coefficient in metals may be estimated approximately from the following expression, valid as $T \to 0$:\cite{Behnia2009}

\begin{equation}
|\frac{\nu}{T}| \simeq \frac{\pi^2}{3} \frac{k_{\rm B}}{e} \frac{\mu}{T_{\rm F}}
\end{equation} 

where $\mu$ is the carrier mobility. (Note that the sign of $\nu$ is difficult to predict.)
This relation is found to hold within a factor two or so for a wide range of metals.\cite{Behnia2009}

Using $T_{\rm F} = 410$~K from quantum oscillations (see above), 
and $\mu = 0.033$~T$^{-1}$ from the Hall angle,\cite{Chang:PRL10}
yields a predicted value of $| \nu/T | = 9~$nV/K$^{2}$~T,\cite{Chang:PRL10}
compared to a value of $\nu/T = - 7~$nV/K$^{2}$~T measured at $T \to 0$ in YBCO at $p = 0.12$.\cite{Chang:PRL10}
This excellent agreement shows that the large, in this case negative, Nernst coefficient observed in the normal state of YBCO at low temperature
is a consequence of the high-mobility electron pocket detected in quantum oscillations. 
This elucidates the hitherto unexplained negative Nernst signal observed in earlier measurements on YBCO in the vicinity of $p = 0.12$.\cite{Ong2004} 

Given that the electron pocket is a small closed pocket, we would expect the Nernst coefficient to be roughly isotropic in the plane, 
with $\nu_a \simeq \nu_b$ at $T \to 0$. 
This is indeed the case, as found recently in measurements up to 28~T, in YBCO at $p=0.12$.\cite{Chang2010}
Therefore, another consequence of the electron pocket forming below $T_{\rm H}$
is to short-circuit the large in-plane anisotropy of the Nernst coefficient that characterizes 
the pseudogap phase at high temperature.\cite{Daou:Nat10}
Specifically, the Nernst anisotropy ratio $(\nu_b - \nu_a)/(\nu_b + \nu_a)$ grows as the temperature decreases from $T^\star$, the pseudogap temperature (see section V.B), 
and then as soon as $R_{\rm H}(T)$ starts to drop, so does $(\nu_b - \nu_a)/(\nu_b + \nu_a)$, 
both falling simultaneously from $T_{\rm max} \simeq 100$~K down to the lowest measured temperature, 
where $(\nu_b - \nu_a)/(\nu_b + \nu_a) \simeq 0$.\cite{Chang2010}


\section{Lifshitz transition}


\subsection{Disappearance of the electron pocket}

In YBCO at $p = 0.078$, $R_{\rm H}$ is positive at all temperatures and simply increases monotonically as $T \to 0$. 
%
We therefore infer that there is no electron pocket in the Fermi surface at that doping. 
Note that quantum oscillations have only been observed at $p > 0.08$, namely in YBCO at $p = 0.09-0.13$,\cite{Doiron:Nat07,Jaudet:PRL08,Sebastian:Nat08,Audouard2009,Singleton2010,Sebastian2010}
and in YBa$_2$Cu$_4$O$_8$,\cite{Yelland:PRL08,Bangura:PRL08}
for which $p \simeq 0.14$, a range of dopings which is strictly contained inside the $T_0$ dome of Fig.~4.
This implies that there must be a topological change in the Fermi surface of YBCO as the doping is decreased below a critical doping  
$p \equiv p_{\rm L}  = 0.08$, a Lifshitz transition at which the closed electron pocket disappears. 
The sudden drop in $T_{\rm max}$ from $\sim 40$~K at $p = 0.083$-0.088 down to zero at $p = 0.078$ (Fig. 4) is one manifestation of this transition.


\begin{figure}[t]
\center
	\includegraphics[width=8cm]{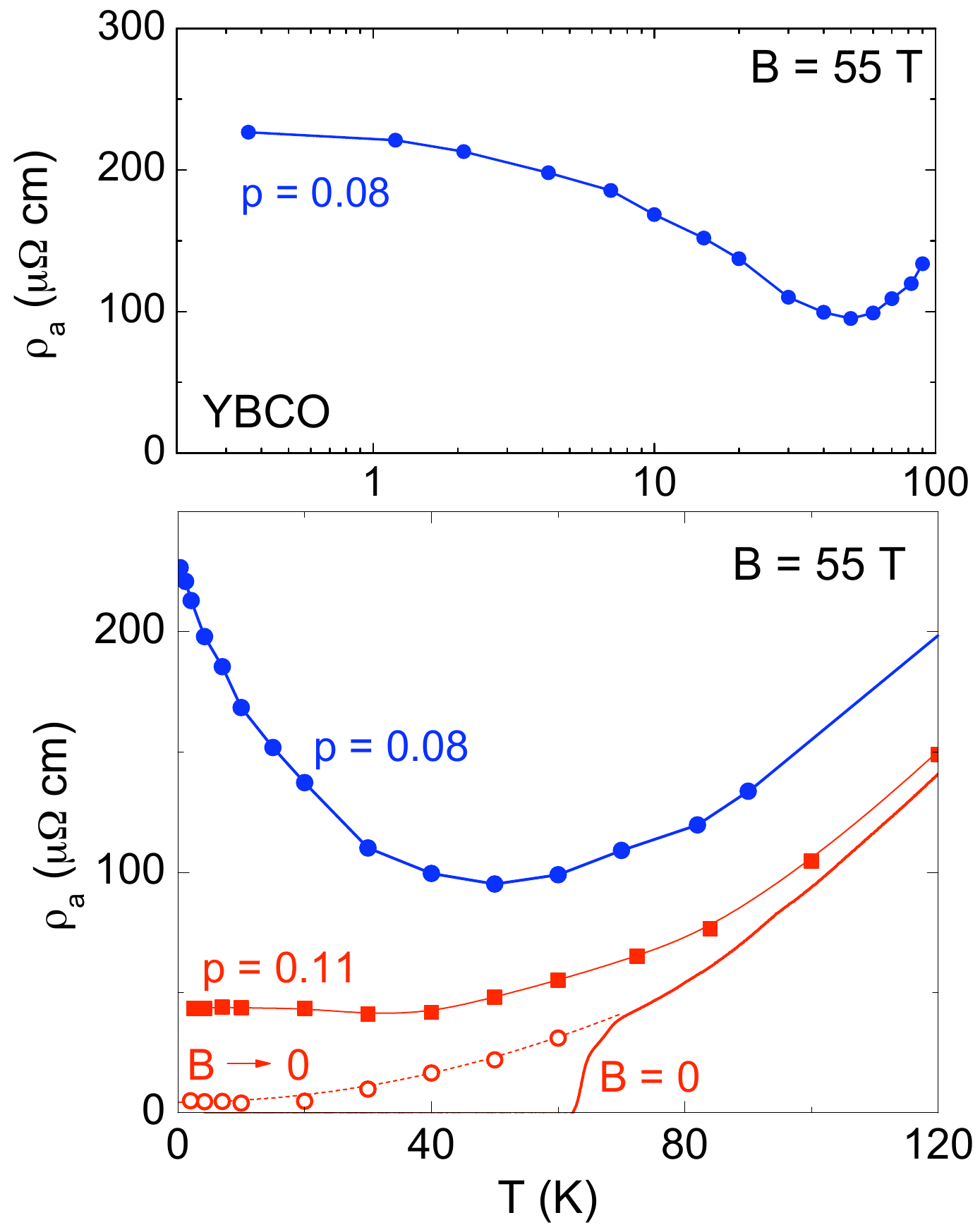}
	\caption{
	Bottom panel:
	In-plane resistivity of YBCO as function of temperature, measured in a magnetic field of 55~T (closed symbols), for $p=0.08$ (blue dots) 
	and $p=0.11$ (red squares).
	The continuous (red) curve is the resistivity in zero field.
	The open (red) circles is the resistivity extrapolated to $B=0$ using a linear fit to the normal-state $\rho(B)$ of Fig.~\ref{fig:rhovsB} and extrapolate
	the zero-field resistivity $\rho(0)$, an estimate of the MR-free normal-state resistivity (see text).
	Top panel:
	data for $p=0.08$ at $B=55$~T plotted on a semi-log plot to zoom on the low-temperature region, where $\rho_a$ is seen to saturate as $T \to 0$,
	a signature of metallic behaviour. 
	}
	\label{fig:rhovsT}
\end{figure}



\begin{figure}[t]
\center
	\includegraphics[width=8cm]{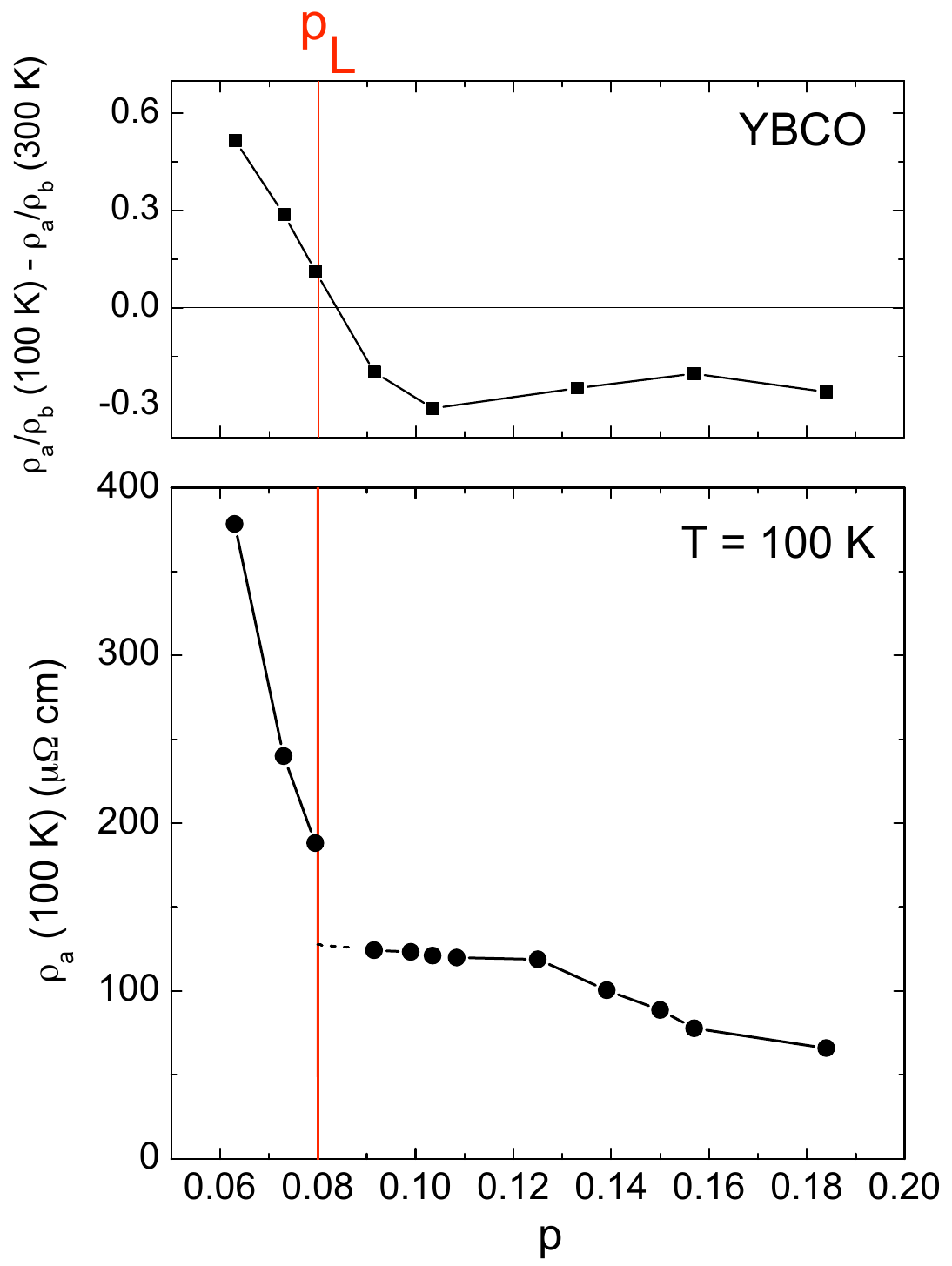}
	\caption{
	Bottom panel:
	Doping evolution of the in-plane zero-field resistivity $\rho_a$ at $T = 100$~K, obtained from the data of ref.~\onlinecite{Ando:PRL04}.
	Top panel:
	Doping evolution of the in-plane anisotropy in the zero-field resistivity, $\rho_a/\rho_b$, 
	measured at $T = 100$~K, relative to its value at 300~K; from ref.~\onlinecite{Sun2004}.
	The vertical (red) line at $p = p_{\rm L} = 0.08$ marks the location of the Lifshitz transition detected in the Hall data (Fig.~4).
	The resistivity is seen to undergo a sudden jump across $p_{\rm L}$, accompanied by an enhancement of the in-plane 
	$a-b$ anisotropy.
	}
	\label{fig:rho100K}
\end{figure}


An immediate consequence of losing the electron pocket is a dramatic drop in magnetoresistance (MR) across $p_{\rm L}$. 
(In the two-band model mentioned above,\cite{Rourke2010} 
the MR would go to zero once the electron pocket disappears.) 
In Fig.~\ref{fig:rhovsB}, we see that the strong MR at $p = 0.11$ becomes very weak at $p \leq 0.08$.
For example, at $T = 60$~K, $(d\rho/dB)(B/\rho)$ at 50~T is 10~\% at $p=0.08$, compared to 100~\% at $p = 0.11$. 
Similarly, the large MR in $\rho_c$ at $p = 0.11$ also disappears when $p$ drops to $\simeq 0.08$.\cite{Balakirev2000}


\subsection{`Metal-insulator' transition}

The disappearance of the high-mobility electron pocket should have a major impact on the conductivity. 
It does.
In Fig.~\ref{fig:rhovsT}, the $a$-axis resistivity at $B = 55$~T of the two YBCO samples with $p = 0.08$ and $p = 0.11$ is plotted as a function of temperature.  
At $p=0.08$, $\rho(B=55$~T) shows a pronounced upturn below 50 K, giving a large value at $T \to 0$. 
At $T = 0.4$~K, $\rho(B=55$~T) $\simeq 225~\mu \Omega$~cm. 
At $p = 0.11$, there is no such upturn and $\rho(B=55$~T) $\simeq 45~\mu \Omega$~cm at $T \to 0$. 

Moreover, because of the large MR at $p=0.11$, $\rho_a$ measured in 55~T overestimates the intrinsic zero-field resistivity of the metal at low temperature, 
by at least a factor 2.
In other words, the MR-free conductivity of YBCO drops by at least a factor 10 in going from $p=0.11$ to $p=0.08$, in the normal state at $T \to 0$. 
A rough estimate of the MR-free resistivity at $T = 10$~K may be obtained by extrapolating the linear MR to $B=0$, as shown by the dotted line in 
Fig.~\ref{fig:rhovsB}. This gives $\rho(0) \simeq 5~\mu\Omega$~cm.
A linear fit to the data of Fig.~\ref{fig:rhovsB} for $p=0.11$ yields the extrapolated zero-field values $\rho(0)$ plotted in Fig.~\ref{fig:rhovsT},
with $\rho(0) \simeq 5-10~\mu\Omega$~cm at $T \to 0$. 
Note that $\rho(0)$ grows approximately as $T^2$, the dependence expected of a Fermi liquid. 

The order-of-magnitude change in $\rho(T \to 0, B \to 0)$ does not occur gradually, but suddenly, at $p \simeq 0.08$. 
This can be seen by examining published data on $\rho_a(T)$ in YBCO as a function of $p$.\cite{Ando:PRL04}
In Fig.~\ref{fig:rho100K}, we plot $\rho_a$ at $T=100$~K for dopings in the range $0.06 < p <  0.19$. 
There is a sudden change in behaviour occurring between 0.08 and 0.09, where $\rho$(100~K) jumps from a low, $p$-independent value above 0.09 to a high, 
strongly $p$-dependent value below 0.08. 

The fact that the Lifshitz transition is accompanied by a 10-fold increase in resistivity is direct evidence that the electron pocket 
has a conductivity roughly 10 times larger than the conductivity of the rest of the Fermi surface, 
thereby explaining why it dominates the Hall and Seebeck coefficients (as discussed in sec.~III.D).
Within a stripe scenario (sec.~V.A.2), the quasi-one-dimensional character of the Fermi surface sheets other than the electron pocket
could explain why they have such a low conductivity. 

The rapid decrease of $\rho(T)$ upon warming from $T=0$ at $p=0.08$ may be related to the rapid increase in $R_{\rm H}(T)$ at $p > p_{\rm L}$, in the following sense:
if the hole-like parts of the Fermi surface are similar on both sides of $p_{\rm L}$, the fact that their conductivity grows rapidly with temperature 
would enhance their positive contribution to the overall Hall signal. 
This is consistent with the findings of the two-band analysis applied to YBa$_2$Cu$_4$O$_8$.\cite{Rourke2010}

Our data clearly links the sudden change in $\rho_a$ across $p_{\rm L}$
to a change in the Fermi-surface topology of YBCO. 
More generally, we propose that the so-called `metal-insulator' crossover in cuprates, studied mostly in La$_{2-x}$Sr$_x$CuO$_4$ (LSCO) and 
Bi$_2$Sr$_{2-x}$La$_x$CuO$_{6+\delta}$ (BSLCO),\cite{Ando1995,Boebinger1996,Ono2000}
is in fact triggered by a change in Fermi-surface topology.
In other words, it is not a metal-insulator transition caused by localization but a Lifshitz transition with an associated loss of conductivity. 
This would explain why the transition occurs at a value of the resistance per plane which is one order of magnitude lower\cite{Ando1995,Boebinger1996,Ono2000}
than expected from the criterion for localization, $k_F l \simeq 1$, where $k_F$ is the Fermi wavevector and $l$ the electronic mean free path. 
Note that the normal-state $\rho_a(T)$ of YBCO at $p < p_{\rm L}$ does not diverge but saturates as $T \to 0$ (see top panel of Fig.~9),
the signature of metallic behaviour.

In addition to that seen at $p_{\rm L}$, a change in the conductivity is also expected at the main
critical  point $p^\star$, where the large Fermi surface of the
overdoped regime first undergoes its reconstruction (in the absence of superconductivity).
This region is not accessible in YBCO, but
measurements in Nd-LSCO, for example, show that $\rho(0)$ increases by a
factor 10 upon crossing $p^\star = 0.235$, going from $\rho_{ab} \simeq 25~\mu\Omega$~cm at $p = 0.24$ 
to $\rho_{ab} \simeq 250~\mu\Omega$~cm at $p = 0.20$.\cite{Daou2009}
In LSCO, the `metal-insulator' crossover becomes apparent as an upturn
in $\rho(T)$ at low temperature when $p$ becomes less than 0.16 or
so.\cite{Ando1995,Boebinger1996}
In other words, the full evolution of the conductivity from
$p > p^\star$ to $p < p_{\rm L}$ will occur in two or more stages as the
Fermi-surface topology undergoes a sequence of changes (as sketched in Fig.~11).

We emphasize that the Lifshitz critical point at $T = 0$ in the non-superconducting state will in general not be located at the same doping 
in the superconducting state, because superconductivity is in general likely to compete with the ordered phase that causes the electron
pocket to emerge (see, for example, ref.~\onlinecite{Sachdev2009}).
As a result, one should not expect changes in the normal-state Fermi surface at $p_{\rm L}$ (or $p^\star$) 
to be reflected in the properties of $d$-wave quasiparticles deep in the superconducting state at the same doping. 
%


\section{Discussion}


\begin{figure}[t]
\center
	\includegraphics[width=8cm]{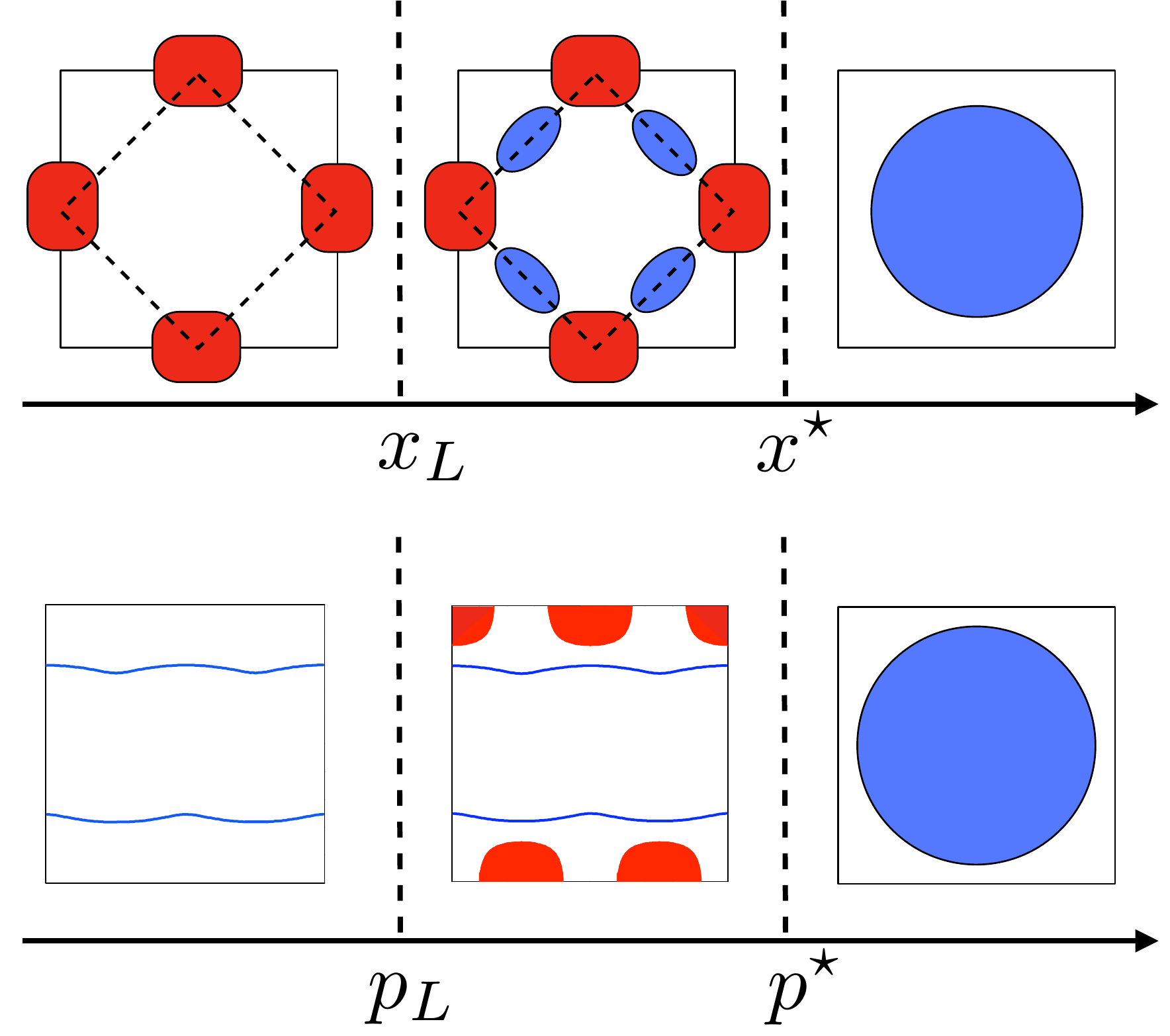}
	\caption{
	Schematic illustration of two scenarios of Fermi-surface reconstruction in cuprates (in the normal state at $T=0$):
	commensurate antiferromagnetic order in electron-doped cuprates (top);\cite{Chubukov1997,Lin2005}
	stripe order in hole-doped cuprates (bottom).\cite{Millis:PRB07}
	$x$ is the electron doping, $p$ the hole doping.
	$x^\star$ and $p^\star$ are the critical dopings for the onset of order (in the absence of superconductivity)
	and the associated breaking of translational symmetry.
	$x_{\rm L}$ and $p_{\rm L}$ are the Lifshitz critical points where the reconstructed Fermi surface undergoes
	a change of topology: a loss of the small hole pocket in the former; a loss of the small electron pocket in the latter.
	Closed blue pockets are hole-like; closed red pockets are electron-like; wavy blue lines are open Fermi surfaces. For $p<p^\star$, only the first quadrant of the square lattice Brillouin zone is shown.
	}
	\label{fig:FScartoons}
\end{figure}

We saw that in YBCO the drop in $R_{\rm H}(T)$ and $S/T$ upon cooling are signatures of the emergence
of an electron pocket in the reconstructed Fermi surface. 
Similar signatures have been observed in several other hole-doped cuprates.
A drop in $R_{\rm H}(T)$ is seen in
LSCO,\cite{Suzuki2002,Balakirev2009} 
BSLCO,\cite{Balakirev2003} 
La$_{2-x}$Ba$_x$CuO$_4$ (LBCO),\cite{Adachi2001,Adachi2010} 
La$_{2-x-y}$Sr$_x$Nd$_y$CuO$_4$ (Nd-LSCO),\cite{Noda1999} 
and 
Eu-LSCO,\cite{Choix:Nat09}
at $p \simeq 0.12$, below $T \simeq 50$-100~K, leading to a negative $R_{\rm H}$ is some cases.\cite{Suzuki2002,Adachi2010}
A lower electron mobility may explain why the otherwise similar drop is shallower in these materials, 
since disorder scattering is thought to be stronger than in YBCO. 
A negative $S/T$ is seen in
LBCO,\cite{Tranquada2007}
Nd-LSCO,\cite{Nakamura1992}
and
Eu-LSCO, at $p \simeq 1/8$.\cite{Chang:PRL10}
We conclude that the emergence of an electron pocket in the Fermi surface at low temperature is not a feature specific to YBCO
(and YBa$_2$Cu$_4$O$_8$) but a general property of underdoped cuprates.
In the next section, we investigate its possible origin.


\subsection{Origin of the electron pocket}

The existence of an electron pocket in the Fermi surface of a hole-doped cuprate implies that the large hole-like Fermi surface of the overdoped regime \cite{Mackenzie1996,Hussey2003,Plate2005,Vignolle:Nat08,Daou2009} must undergo a fundamental transformation at a critical doping between $p \simeq 0.25$ and $p=0.15$, 
in the normal state at $T=0$ (once superconductivity is suppressed). 
A natural mechanism for such a transformation is a Fermi-surface reconstruction caused by the onset of some new periodicity due to an ordered phase
that breaks the translational symmetry of the lattice.\cite{Chakravarty2008}
Some of the possible ordered phases include:
commensurate antiferromagnetic order (AF),\cite{Chubukov1997}
commensurate $d$-density-wave order (dDW),\cite{Chakravarty:PNAS08}
incommensurate spin-density-wave order (SDW),\cite{Harrison:PRL09}
and stripe order,\cite{Millis:PRB07}
a form of unidirectional SDW with an associated charge-density wave.\cite{Kivelson2003,Vojta2009}

In this section, we first sketch the AF and stripe scenarios and outline how the former applies to the electron-doped cuprates
Pr$_{2-x}$Ce$_x$CuO$_4$ (PCCO) and Nd$_{2-x}$Ce$_x$CuO$_4$ (NCCO) 
and the latter applies to the hole-doped cuprates Nd-LSCO and Eu-LSCO.
We then examine which of these scenarios of Fermi-surface reconstruction might apply to YBCO.


\subsubsection{Fermi-surface reconstruction by commensurate antiferromagnetic order in electron-doped cuprates}

Hall effect measurements in the electron-doped cuprate PCCO showed that the large hole-like Fermi surface of the overdoped regime
is reconstructed below a critical electron doping $x \equiv x^\star = 0.165 \pm 0.005$.\cite{Dagan2004}
In the $T=0$ limit, $R_{\rm H}$ changes sign across $x^\star$, going from $+ 1/(1 - x)$ above to
$- 1/x$ well below $x^\star$.  
The Seebeck coefficient exhibits the same evolution, with $S/T$ at $T \to 0$ going from small and positive at $x > x^\star$,
to large and negative at $x < x^\star$.\cite{Li2007} 
The reconstructed Fermi surface must therefore contain a small electron pocket.

The Hall data is well described by a scenario of band folding by an antiferromagnetic (AF) order with $(\pi, \pi)$ wavevector.\cite{Lin2005}
This scenario predicts a two-stage evolution of the Fermi surface, shown schematically in Fig.~\ref{fig:FScartoons} (top panel).
Just below $x^\star$, the Fermi surface is expected to contain both an electron pocket around $(\pm \pi, 0)$ and $(0, \pm \pi)$ 
and a smaller hole pocket around $(\pm \pi/2, \pm \pi/2)$.  
As the AF potential grows with $x$ decreasing away from $x^\star$, the small hole pocket eventually disappears, at a Lifshitz critical point
$x_{\rm L}$.\cite{Lin2005}
Because experimentally $R_{\rm H}(T \to 0)$ is negative just below $x^\star$, there is no sign change in 
$R_{\rm H}(T \to 0)$ across $x_{\rm L}$ and it is therefore difficult to pin down the Lifshitz critical point from 
Hall measurements.

ARPES measurements\cite{Armitage2001,Matsui2005,Matsui2007} on the closely-related material NCCO have seen the three Fermi-surface topologies
sketched in Fig.~\ref{fig:FScartoons}, and are therefore fully consistent with the AF scenario.
Quantum oscillations have also detected a large Fermi surface above $x^\star$ and a small pocket below $x^\star$, 
attributed to the hole pocket.\cite{Helm:PRL09}
Long-range static antiferromagnetic order, with a commensurate wavevector $Q = (\pi, \pi)$, was observed in NCCO with neutron scattering 
from $x = 0$ up to $x \simeq 0.13$.\cite{Motoyama2007}
The fact that AF order appears to end at the onset of superconductivity (at $x \simeq 0.13$) whereas the normal-state quantum critical point seen in 
the Hall effect once superconductivity is suppressed with a magnetic field is located at $x = 0.165$ may have to do with the competing effect
of superconductivity, which pushes AF order (in zero field) to lower doping.\cite{Sachdev2009}

All in all, there is little doubt that the Fermi surface of electron-doped cuprates undergoes a two-stage reconstruction 
at $x^\star$ and $x_{\rm L}$ due to commensurate AF order.\cite{Armitage2010}
This Fermi-surface reconstruction is accompanied by an increase in the resistivity, just as it does in YBCO and Nd-LSCO. 
The in-plane resistivity $\rho_{a}(T)$ of PCCO goes from metallic-like at $x > x^\star$, with a low value at $T \to 0$,
to an insulating-like resistivity at $x < x^\star$, with an upturn at low temperature.\cite{Dagan2004} 
The resistivity at $T \to 0$ goes from $\rho_{0} \simeq 25~\mu\Omega$~cm at $x = 0.16$
to $\rho_{0} \simeq 175~\mu\Omega$~cm at $x = 0.13$,\cite{Dagan2004}
a seven-fold increase very similar to that observed in YBCO going from $p = 0.11$ to $p = 0.08$ (Fig.~8).




\subsubsection{Fermi-surface reconstruction by stripe order in hole-doped cuprates}

In the hole-doped cuprates Nd-LSCO and Eu-LSCO, a form of slowly fluctuating combined spin-charge density wave called `stripe order'
is detected by various experimental probes, including:
neutron diffraction,\cite{Tranquada1995,Ichikawa2000}
x-ray diffration,\cite{Zimmermann1998,Niemoller1999,Fink2009}  
nuclear quadrupole resonance,\cite{Hunt2001}
and muon spin relaxation.\cite{Nachumi1998}
The wavevector of the spin order is incommensurate, with $Q = (\pi, \pi + \delta)$.

This stripe order sets in below a temperature which depends on the time scale of the probe.
For example, x-ray diffraction yields an onset temperature
$T_{\rm CO} = 80 \pm 10$~K in Eu-LSCO at $p = 0.125$.\cite{Fink2009,Choix:Nat09}
Above $p = 0.125$, the onset temperature decreases monotonically with doping, and vanishes at
$p \simeq 0.25$.\cite{Ichikawa2000,Hunt2001}

Transport measurements in Nd-LSCO show major changes across the quantum critical point where stripe order vanishes,
and these can be used to pin down the location of the critical doping where Fermi-surface reconstruction occurs,
at $p^\star = 0.235 \pm 0.005$.\cite{Daou2009,Cyr-Choiniere2010}
The Hall coefficient goes from a small positive value at $p = 0.24$, where $R_ {\rm H}(T \to 0) = + 1 / (1 + p)$, 
consistent with a single large hole-like Fermi surface in the overdoped regime, 
to a large positive value at $p = 0.20$.\cite{Daou2009}
As mentioned already (in sec.~IV.B), the in-plane normal-state resistivity $\rho_{ab}$ of Nd-LSCO at $T \to 0$ undergoes a ten-fold increase between 
$p = 0.24$ and $p = 0.20$.\cite{Daou2009} 
Reminiscent of the `metal-insulator' transition across $p_{\rm L}$ in YBCO, it is also 
a metal-to-metal transition, this time across $p^\star$.
It was tracked in detail in measurements of $\rho_c$.\cite{Cyr-Choiniere2010}

Upon cooling, the Hall coefficient of Eu-LSCO at $p = 0.125$ undergoes a pronounced decrease 
below $\sim 100$~K,\cite{Choix:Nat09}. 
The onset of this drop is not sudden but smooth, and it coincides with the onset of
stripe order detected by x-ray diffraction.
This is compelling evidence that stripe order causes a Fermi-surface reconstruction 
in Eu-LSCO.\cite{Taillefer:LT25}  
In Eu-LSCO, the drop in $R_{\rm H}(T)$ is not quite deep enough to make $R_{\rm H}$ go negative.
However, the related drop in the Seebeck coefficient $S$ does make $S/T$ go deeply negative at $T \to 0$.\cite{Chang:PRL10}
Similar drops are observed in $R_{\rm H}$ and $S/T$ at $p \simeq 0.12$ in the other two hole-doped cuprates that show clear signatures of stripe order,
namely LBCO\cite{Adachi2001,Adachi2010,Tranquada2007}
and Nd-LSCO,\cite{Nakamura1992,Noda1999}
again coincident with the onset of stripe order, at comparable temperatures (between 50 and 80~K).

Hall measurements also reveal a drop in $R_{\rm H}(T)$ below $T \simeq 60 - 100~$K in underdoped
BSLCO\cite{Balakirev2003} and LSCO.\cite{Suzuki2002,Balakirev2009}
Neutron diffraction shows static SDW order in LSCO with the same wavevector (same incommensurability $\delta$) as in Nd-LSCO,\cite{Chang2008,Chang2009}
up to $p \simeq 0.13$ in $B=0$ and up to higher doping in applied magnetic fields.\cite{Chang2008,Chang2009,Khaykovich2005}
At $p=0.12$, NMR/NQR reveal identical signatures of stripe order in LSCO as in Nd-LSCO and Eu-LSCO.\cite{Curro2000,Hunt2001,Julien2001,Mitrovic2008}
This strongly suggests that a form of stripe order very similar to that found in Nd-LSCO (and Eu-LSCO and LBCO) is responsible for a 
Fermi-surface reconstruction in LSCO.

Calculations show that stripe order causes a reconstruction of the large hole-like Fermi surface into electron pockets and quasi-1D sheets,
as sketched in Fig.~11, with the additional possibility of some small hole pockets.\cite{Millis:PRB07}
The electron pocket is expected to disappear with underdoping, at a Lifshitz transition, 
either because the density-wave potential gets stronger,\cite{Lin2008}
in which case the pocket gradually shrinks to nothing, 
or because the incommensurability $\delta$ gets smaller,\cite{Norman2010}
in which case the pockets get closer and closer until they coalesce to form a new quasi-1D open Fermi surface
(not drawn in Fig.~11).
(In LSCO, $\delta$ is known experimentally to decrease with underdoping below $p \simeq 1/8$.\cite{Yamada1998})


\subsubsection{Fermi-surface reconstruction in YBCO}

The AF scenario that applies to electron-doped cuprates could in principle also apply to hole-doped cuprates, with the same 
Fermi-surface topology at high doping (large hole-like surface above $p^\star$) 
and intermediate doping (small electron and hole pockets between $p_{\rm L}$ and $p^\star$),
but with a different topology at low doping, where instead of having only a small electron pocket at $(\pm \pi, 0)$ and
$(0, \pm \pi)$
the Fermi surface below $p_{\rm L}$ would contain only small hole pockets at $(\pm \pi/2, \pm \pi/2)$ (see Fig.~11). 
The stripe scenario could also in principle apply to YBCO, as could the incommensurate SDW scenario.

The first question is what type of order is actually observed in the relevant doping range.
Commensurate AF order is observed up to $p \simeq 0.05$, namely up to where the superconducting phase begins (on the underdoped side).
Static incommensurate SDW order has been observed in YBCO with neutron scattering at $p \simeq 0.07$ ($T_c = 35$~K).\cite{Hinkov2008,Haug2009}
At $p > 0.07$, no static SDW order is detected in zero field.
But this could be due to the competing effect of the superconducting phase.\cite{Sachdev2009}
Whether SDW order would persist to higher doping in the presence of a large field that suppresses superconductivity, 
as it does in the case of LSCO,\cite{Chang2008,Chang2009}
is not known yet for the case of YBCO, 
although a magnetic field does enhance the SDW moment at $p \simeq 0.07$.\cite{Haug2009}

The reported SDW order not only breaks the translational symmetry but also the four-fold rotational symmetry of the CuO$_2$ planes.\cite{Hinkov2008} 
We would therefore expect the Fermi-surface reconstruction caused by this SDW order to result in a Fermi surface with some two-fold in-plane anisotropy. 
Indeed, one way to distinguish between the AF (or dDW) scenario and the stripe (or unidirectional SDW) scenario, is to look for evidence of
broken rotational symmetry. Here we mention two instances of such evidence.

%
First, the in-plane resistivity of YBCO at $p < 0.08$ is increasingly anisotropic at low temperature, 
reaching an anisotropy ratio $\rho_a/\rho_b \simeq 2$.\cite{Ando:PRL02} 
This means that in the metallic state below the critical doping $p_{\rm L} = 0.08$, the Fermi surface that remains after the electron pocket has disappeared must have two-fold anisotropy in the plane. 

The presence of the high-mobility electron pocket above $p_{\rm L}$ will short-circuit such in-plane anisotropy in the resistivity, 
just as it short-circuits the anisotropy in the Nernst signal at low temperature (see sec.~III.E.2).
This is precisely what is seen in published data on the doping dependence of the anisotropy ratio,\cite{Ando:PRL02}
reproduced in the top panel of Fig.~10:
$\rho_a/\rho_b$ jumps as $p$ drops below $p_{\rm L}$.
Knowing now that a high-mobility electron pocket appears suddenly in the Fermi surface of YBCO above $p=0.08$ 
explains why a large $\rho_a/\rho_b$ is not seen near $p = 1/8$, where stripe order is normally expected to be most stable.

A second evidence that the reconstructed Fermi surface of YBCO is anisotropic comes from measurements of the microwave conductivity $\sigma_1$ in the superconducting state at low temperature.\cite{Harris2006} 
Whereas $\sigma_1$ is perfectly isotropic at high doping, with $\sigma_{1b}/\sigma_{1a} = 1.0$ below 30 K at $p=0.18$, it is strongly anisotropic at low doping,  
with $\sigma_{1b}/\sigma_{1a} = 3.5$ below 30 K at $p=0.10$.\cite{Harris2006}
(Note that CuO chains in the structure of YBCO cannot be responsible for this anisotropy because charge carriers in the highly imperfect quasi-1D chains 
localize at low temperature, especially in the underdoped regime.\cite{Ando:PRL02} 
Note also that chains conduct much better at $p=0.18$, where $\sigma_{b}/\sigma_{a} = 4.5$ at 100 K,\cite{Daou:Nat10}
than they do at $p=0.10$, where $\sigma_{b}/\sigma_{a} \simeq 1.0$ at 100 K.\cite{Doiron:Nat07}) 
This implies that in the superconducting state at $B=0$ the Fermi surface of YBCO (associated with the CuO$_2$ planes) 
undergoes a transformation that breaks its original rotational symmetry. 
The critical doping for this transformation, which we identify with the reconstruction of the large hole-like Fermi surface of the overdoped regime, 
is somewhere between $p=0.10$ and $p=0.18$. 
It is most likely shifted down in doping relative to the normal-state critical point $p^\star$, as expected from the phase competition between SDW order and superconductivity.\cite{Sachdev2009}




Further support for a stripe scenario being relevant to YBCO comes from a detailed comparison between YBCO and Eu-LSCO.
%
It has been shown \cite{Taillefer:LT25} that the Hall coefficient drops in identical fashion in YBCO and Eu-LSCO at $p=0.12$, at the very same temperature for the same doping.
The same is true of the Seebeck coefficient, with $S/T$ dropping in very similar fashion for both materials at that same doping, 
changing sign from positive to negative at the same temperature.\cite{Chang:PRL10}
It is difficult to imagine that two different mechanisms would be responsible for such similar transport signatures of Fermi-surface reconstruction.

In summary, a stripe scenario appears quite reasonable for YBCO:
it can not only account for the appearance of an electron pocket in the Fermi surface below $p^\star$, a result of broken translational symmetry, 
and for its disappearance at $p_{\rm L}$, due to a Lifshitz transition, 
it is also consistent with the evidence of broken rotational symmetry from neutron scattering,\cite{Hinkov2008} 
resistivity,\cite{Ando:PRL02} 
and microwave conductivity.\cite{Harris2006} 
Of course, this assumes that stripe (or unidirectional SDW) order persists to dopings of order $p \simeq 0.2$ in the absence of superconductivity.

Whether the Lifshitz transition at $p_{\rm L}$ is due to an increase in the stripe potential or a decrease in the incommensurability $\delta$
remains to be seen. 
Detailed measurements of the quantum oscillations in YBCO as $p_{\rm L}$ is approached from above could help discriminate between the two mechanisms. 
It has been argued\cite{Norman2010} that the recently-reported increase in the cyclotron mass $m^\star$ as $p \to 0.08$ from above\cite{Sebastian2010}
is consistent with a decrease in $\delta$.


 
 \begin{figure}[t]
\center
	\includegraphics[width=8cm]{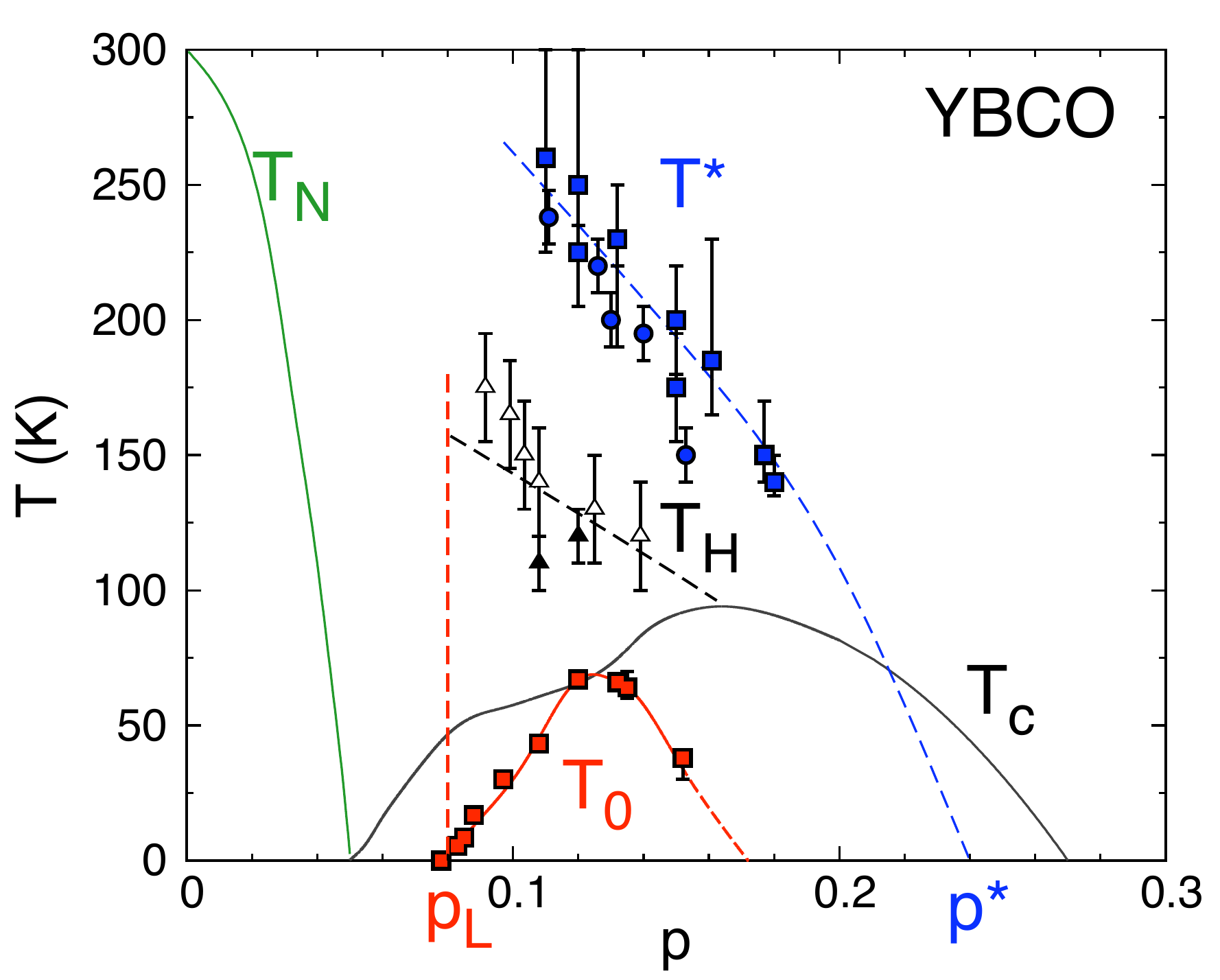}
	\caption{
	Phase diagram of YBCO showing the superconducting phase below $T_c$ (solid black line) and the antiferromagnetic phase below $T_{\rm N}$ (solid green line),
	both in zero magnetic field.
	The lines marked $T_0$ (red squares and red line) and $T_{\rm H}$ (triangles and black dashed line) are taken from Fig.~4.
	The dashed blue line marked $T^\star$ is the boundary of the pseudogap phase, defined as the temperature below which the resistivity (blue circles) and the Nernst 
	coefficient (blue squares) deviate from their high-temperature behaviour (taken from ref.~\onlinecite{Daou:Nat10}). 
	The end of the $T^\star$ line defines the quantum critical point where the large hole-like Fermi surface of the overdoped regime undergoes 
	a reconstruction (in the absence of superconductivity). 
	Here it is taken to be at $p \simeq 0.24$, the same location as in Nd-LSCO.\cite{Daou2009,Cyr-Choiniere2010}
	The red vertical dashed line marks the Lifshitz critical point $p_{\rm L}$.  
	All other dashed lines are a guide to the eye.
	}
	\label{fig:tstar}
\end{figure}
 

\subsection{Relation to the pseudogap phase}

If the first sign of a Fermi-surface modification is detectable in the Hall effect at a temperature $T_{\rm H}$, at what temperature $T_{\rho}$ is it detectable in the resistivity $\rho(T)$? 
At $p=0.12$, where $T_{\rm H} \simeq 120$~K (see Fig.~\ref{fig:T_0}),
$\rho_a(T)$ is linear in temperature at high temperature and it starts to drop below $T_{\rho} \simeq 220$~K.\cite{Ando:PRL04}
The onset of this drop is a standard definition of the temperature $T^\star$ which delineates the pseudogap phase.\cite{Ito1993}
It was recently found that the quasiparticle Nernst coefficient $\nu(T)$ also starts deviating from its high-temperature behaviour at the same temperature, 
so that $T_{\nu} \simeq T_{\rho} \equiv T^\star$.\cite{Matusiak2009,Daou:Nat10}
In Fig.~\ref{fig:tstar}, a phase diagram shows how $T_{\nu}$ and $T_{\rho}$ evolve with $p$ in YBCO, tracing out the pseudogap phase boundary. In the doping range where $R_{\rm H}$ shows a drop, and hence $T_{\rm H}$ can be defined, we see that $T^\star  \simeq 2~T_{\rm H}$, as noted previously.\cite{Abe1999,Xu1999} 

That the resistivity and the Nernst coefficient detect changes in electronic behaviour before they are visible in the Hall or Seebeck coefficients may have to do with the fact that the former two transport coefficients are directly sensitive to the scattering rate whereas the latter two are not, at least in the case of a single Fermi surface. So the full process of reconstructing the Fermi surface, as it unfolds with decreasing temperature, may start at $ T^\star$ with a $k$-dependent change in the scattering before it gives birth to an electron pocket below $T_{\rm H}$, produced by a folding of the Brillouin zone due to the broken translational symmetry. 

A similar two-stage process is observed in Nd-LSCO and Eu-LSCO, whereby $T_{\nu} = T_{\rho} \simeq 2~T_{\rm CO}$,\cite{Choix:Nat09,Taillefer:LT25,Taillefer2010}
where $T_{\rm CO}$ is the onset temperature for stripe order as detected by either x-ray diffraction or nuclear quadrupolar resonance, which coincides with the anomalies in the Hall and Seebeck coefficients 
.\cite{Daou2009, Choix:Nat09,Taillefer:LT25,Chang2010}
This suggests that the pseudogap phase below $T^\star$ is a high-temperature precursor of the stripe order that sets in at a lower temperature. Measurements of the Nernst effect in YBCO support this interpretation in the sense that the Nernst coefficient $\nu(T)$ acquires a strong in-plane anisotropy upon entry into the pseudogap phase, starting at $T_\nu = T^\star$ and rising with decreasing temperature, reaching values as high as $\nu_b/\nu_a \simeq 7$ at $p=0.12$.\cite{Daou:Nat10}
%
%
An interpretation of the pseudogap phase in hole-doped cuprates as a precursor region of stripe (or SDW) fluctuations is analogous to the interpetation of the pseudogap
phase in electron-doped cuprates,\cite{Kyung2004}
where $T^\star$ is found to be the temperature below which the antiferromagnetic correlation length exceeds the thermal de Broglie 
wavelength of the charge carriers.\cite{Motoyama2007}



 \begin{figure}[t]
\center
	\includegraphics[width=8cm]{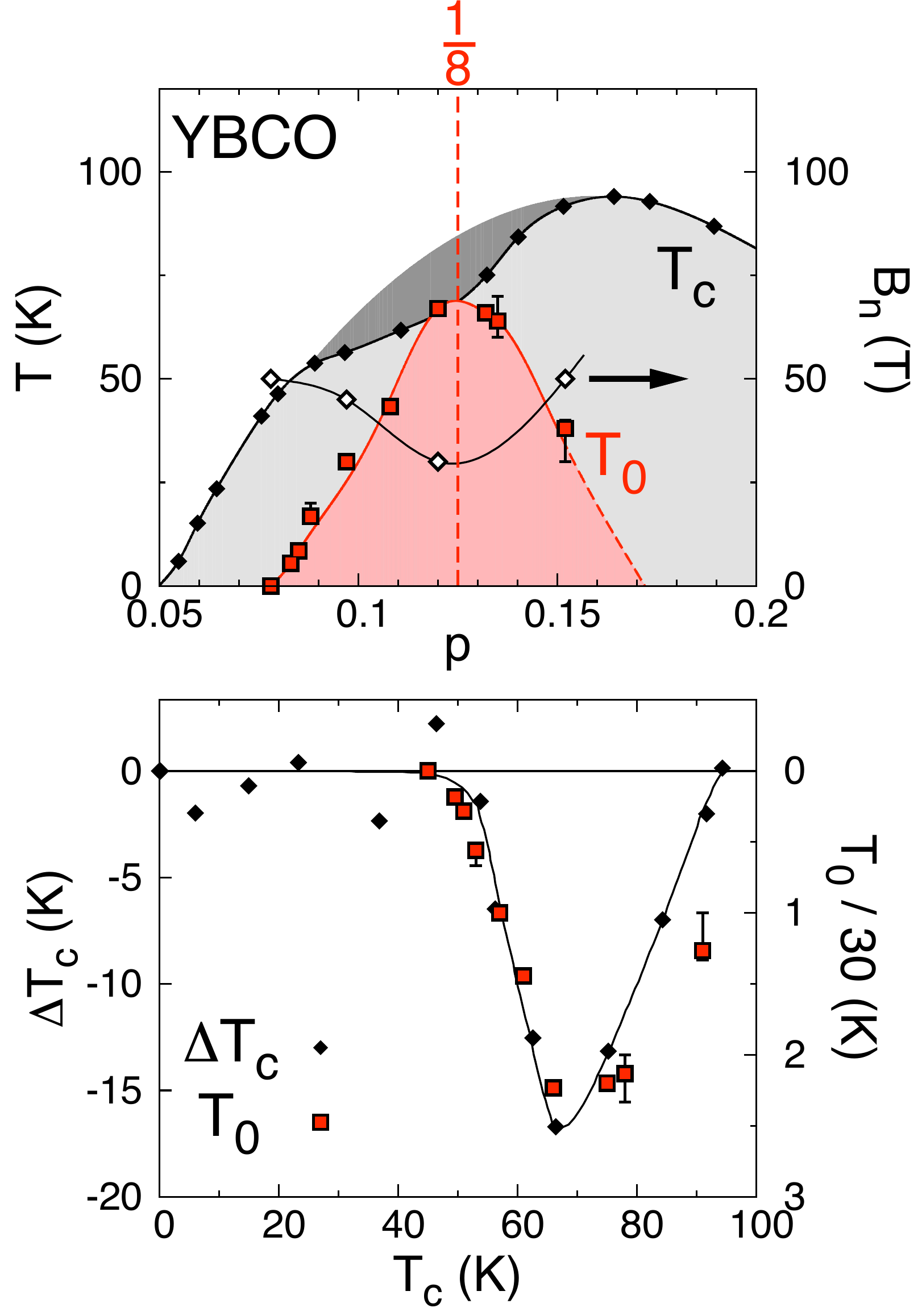}
	\caption{
	Top panel: Phase diagram of YBCO, showing the zero-field superconducting transition temperature $T_c$ (black diamonds) as a function of doping 
	(from ref.~\onlinecite{Liang:PRB06}) and the resistive upper critical field $B_n$ at $T = 10$~K (white diamonds), 
	marking the end of the superconducting transition as a function of field (see Figs.~1, 5 and 6). 
	The dark grey region shows the missing area relative to an idealized parabolic variation of $T_c$ vs $p$.
	The sign-change temperature $T_0$, obtained once superconductivity is suppressed by a magnetic field, 
	is also shown (red squares). 
	Bottom panel: Comparison of the dip in $T_c$, $\Delta T_c$ (black diamonds, left axis; from ref.~\onlinecite{Liang:PRB06}), 
	defined as the suppression of $T_c$ relative to its idealized parabolic dependence (see top panel), 
	with the sign-change temperature $T_0$ (red squares, right axis; from Fig.~4). 
	All lines are a guide to the eye.}
	\label{fig:deltatc}
\end{figure}


\subsection{Relation to superconductivity}

It is interesting to consider whether the evolution of the Fermi surface has any impact on the superconductivity.
A first observation is that $T_{\rm H}(p)$, which we may regard as the onset of Fermi-surface reconstruction,
crosses the $T_c(p)$ line where $T_c$ is maximal (Fig.~\ref{fig:tstar}).
In other words, in the overdoped region, $T_c$ rises with underdoping up until the Fermi surface of the normal state, 
out of which superconductivity emerges, reconstructs.
Superconductivity appears to be weakened by this reconstruction.

It is well known that in hole-doped cuprates there is an anomalous weakening of superconductivity near $p = 1/8$, whereby $T_c$ drops below the idealized parabolic dependence on doping. This is shown in Fig.~\ref{fig:deltatc} for YBCO. 
The difference between the measured $T_c$ and the idealized $T_c$, $\Delta T_c \equiv T_c - T_c^0$, is plotted in Fig.~\ref{fig:deltatc} 
(from ref.~\onlinecite{Liang:PRB06}). 
We see that $\Delta T_c$, the amount by which $T_c$ is suppressed, is largest at $p = 1/8$ ($T_c = 69$~K).
This $T_c$ dip at $p = 1/8$, also seen in LSCO, LBCO, and Nd-LSCO is attributed to the competing effect of stripe order, made more stable 
by a commensurate locking with the lattice when $p = \delta$.
This is compelling evidence that stripe order, or at least stripe correlations, are present in YBCO.
Similarly, the fact that $T_0$ - a marker of Fermi-surface reconstruction - peaks at $p = 1/8$ (Fig.~\ref{fig:deltatc}) also suggests that Fermi-surface reconstruction
is caused by stripe order.
It is intriguing that $\Delta T_c$ is directly proportional to $T_0$ (see bottom panel of Fig.~13). 
Below $p = 0.12$, $\Delta T_c$ and $T_0$ scale perfectly, both vanishing as $p \to p_{\rm L}$. 
It appears as though $T_c$ is suppressed by a more prominent electron pocket.

While the zero-field $T_c$ vs $p$ curve shows a small depression at $p = 1/8$, producing the well-known 60-K plateau,
the resistive upper critical field $B_{n}$ shows a pronounced minimum at $p=1/8$.
From Fig.~\ref{fig:RHvsB}, we see that $B_{n} \simeq 50$~T at $p=0.078$, larger than it is at $p=0.12$, where $B_{n} \simeq 30$~T. 
This is surprising since $T_c$ is considerably lower at $p=0.078$ (45~K) than at $p=0.12$ (66~K). 
Above $p=1/8$, $B_{n}$ rises again, with $B_{n} \simeq 50$~T at $p=0.152$ and $B_{n} > 60$~T at $p=0.16$ (see Fig.~13).
This again points to stripe order.

It remains to be seen whether the onset of superconductivity is directly affected by changes in the Fermi surface or whether both changes are instead driven by an underlying competition between stripe order, say, and $d$-wave superconducting order, with the former being possibly stabilized at $p \simeq 1/8$ by a commensurate locking of the spin/charge modulation with the lattice.


\section{Summary}

Measurements of the Hall coefficient $R_{\rm H}$ in YBCO performed in magnetic fields large enough to suppress 
superconductivity reveal that the normal-state $R_{\rm H}$ is negative at $T \to 0$ in a range of doping from 
$p = 0.083$ to at least $p = 0.152$, in the underdoped region of the phase diagram.
The negative $R_{\rm H}$ is attributed to the presence of a high-mobility electron pocket in the Fermi surface 
of underdoped YBCO at low temperature, located at the ($\pi, 0$) point in the Brillouin zone and responsible for the
quantum oscillations observed in the in-plane transport,\cite{Doiron:Nat07} 
the magnetization\cite{Jaudet:PRL08,Sebastian:Nat08} 
and the $c$-axis resistivity.\cite{Ramshaw2010} 

We attribute the sudden change in $R_{\rm H}$ at $T \to 0$ from negative to positive as the doping is reduced below 
$p = 0.08$ to a change in Fermi-surface topology, or Lifshitz transition,
whereby the electron pocket disappears below a critical doping $p_{\rm L} = 0.08$.
The loss of the high-mobility electron pocket explains a number of previous observations, including:
1) the sudden increase in resistivity across $p_{\rm L}$,\cite{Ando:PRL04}
referred to as a `metal-insulator' transition;
2) the qualitative change in the $c$-axis resistivity across $p_{\rm L}$;\cite{Balakirev2000}
3) the sudden increase in the in-plane anisotropy $\rho_a/\rho_b$ across $p_{\rm L}$.\cite{Ando:PRL02}

The natural explanation for an electron pocket at $(\pi, 0)$ is a Fermi-surface reconstruction caused by an ordered state 
that breaks translational symmetry.
From the empirical fact that similar signatures of an electron pocket are observed in the Hall and Seebeck coefficients of a 
number of hole-doped cuprates, in particular some where stripe order is clearly the cause of the Fermi-surface 
reconstruction,\cite{Taillefer:LT25,Chang:PRL10,Taillefer2010}
we infer that a similar type of stripe order, or undirectional spin-density-wave order, is also at play in YBCO. 

A model of Fermi-surface reconstruction by stripe order\cite{Millis:PRB07} can account not only for the presence of an electron pocket at $(\pi, 0)$
but also for its loss across a Lifshitz transition,\cite{Norman2010} 
a natural consequence of increasing either the stripe potential or the stripe period.
The Fermi surface that remains is made of quasi-1D open sheets, from which a strong in-plane anisotropy of transport is expected, as seen.


\begin{center}
{\bf Acknowledgements}
\end{center}

The authors would like to thank A. J. Millis, M. R. Norman and S. Sachdev for fruitful discussions, 
H. Zhang for help with the sample preparation, and J. Corbin for help with the experiments.
R. L., D. A. B., and W. N. H. acknowledge support from NSERC. 
B. V. and C. P. acknowledge support from the French ANR DELICE.
L. T. acknowledges support from the Canadian
Institute for Advanced Research, a Canada Research Chair,
NSERC, CFI, and FQRNT. Part of this work
has been supported by FP7 I3 EuroMagNET II.



\end{document}